\documentclass[a4paper]{PoS}

\usepackage{tikz}
\usetikzlibrary{matrix,cd,arrows}

\usepackage{amsmath}
\usepackage{amssymb,amscd}
\usepackage{amsfonts}
\usepackage{mathrsfs}
\usepackage{mathtools}
\usepackage{axodraw2}

\usepackage{bbm}

\newcommand{\makecommand}[3]{%
    \foreach \i in #3 {%
        \expandafter\xdef\csname #1\i\endcsname{\noexpand#2{\unexpanded\expandafter{\i}}}%
    }%
}
\makecommand{fr}{\mathfrak}{{a,c,der,g,h,A,X,gl,sl,o,so,osp,u,su,sp,spin,string,Poisson}}
\makecommand{sf}{\mathsf}{{p,e,h,id,B,String,Lie,Hom,SU,Diff,HSymp,L,m,T,Sym,P,E,H,D,pr,M}}
\makecommand{rm}{\mathrm}{{d,e,i,Ham,id}}
\makecommand{ca}{\mathcal}{{A,L,O,Q,G}}
\makecommand{I}{\mathbbm}{{N,Z,R,C}}
\newcommand{\eps}{\varepsilon}
\newcommand{\ewith}{~~~\mbox{with}~~~}
\newcommand{\eand}{~~~\mbox{and}~~~}
\newcommand{\rint}{{\rm int}}

\title{Perturbative Quantum Field Theory\\ and Homotopy Algebras}

\ShortTitle{Perturbative Quantum Field Theory and Homotopy Algebras}

\author{Branislav Jur{\v c}o\\
    Faculty of Mathematics and Physics, Mathematical Institute\\ 
    Charles University Prague, Prague 186 75, Czech Republic\\
    E-mail: \email{branislav.jurco@gmail.com}
}

\author{Hyungrok Kim\\
    Maxwell Institute and Department of Mathematics\\
    Heriot--Watt University, Edinburgh EH14 4AS, United Kingdom\\
    E-mail: \email{hk55@hw.ac.uk}
}

\author{Tommaso Macrelli\\
    Department of Mathematics, University of Surrey\\ Guildford GU2 7XH, United Kingdom\\
    E-mail: \email{t.macrelli@surrey.ac.uk}
}

\author{\speaker{Christian Saemann}\thanks{Report numbers: DMUS--MP-20/01, EMPG--20--06}\\
    Maxwell Institute and Department of Mathematics\\
    Heriot--Watt University, Edinburgh EH14 4AS, United Kingdom\\
    E-mail: \email{c.saemann@hw.ac.uk}
}

\author{Martin Wolf\\
    Department of Mathematics, University of Surrey\\ Guildford GU2 7XH, United Kingdom
    E-mail: \email{m.wolf@surrey.ac.uk}
}

\abstract{We review the homotopy algebraic perspective on perturbative quantum field theory: classical field theories correspond to homotopy algebras such as $A_\infty$- and $L_\infty$-algebras. Furthermore, their scattering amplitudes are encoded in minimal models of these homotopy algebras at tree level and their quantum relatives at loop level. The translation between Lagrangian field theories and homotopy algebras is provided by the Batalin--Vilkovisky formalism. The minimal models are computed recursively using the homological perturbation lemma, which induces useful recursion relations for the computation of scattering amplitudes. After explaining how the homolcogical perturbation lemma produces the usual Feynman diagram expansion, we use our techniques to verify an identity for the Berends--Giele currents which implies the Kleiss--Kuijf relations.}

\FullConference{Corfu Summer Institute 2019 "School and Workshops on Elementary Particle Physics and Gravity" (CORFU2019)\\
    31 August - 25 September 2019\\
    Corfu, Greece}

\begin{document}
    
    \section{Introduction}
    
    Homotopy algebras such as $A_\infty$- or $L_\infty$-algebras arise very naturally in string theory. For example, the $B$-field in string theory is part of the connective structure of a gerbe, a higher or categorified versions of a principal circle bundle. Such a higher bundle  comes naturally with higher versions of symmetries, which are most conveniently encoded in $L_\infty$-algebras. Similarly, the intricate gauge symmetries arising from the underlying diffeomorphism invariance of string theory fixes the actions of open and closed string field theories to be the homotopy Maurer--Cartan theories of $A_\infty$- and $L_\infty$-algebras, respectively.
    
    More surprising may be the fact that homotopy algebras arise equally naturally already in the context of ordinary quantum field theory and, moreover, that they are very useful in the analysis of correlation functions and scattering amplitudes. Indeed, they capture literally the usual description of tree level amplitudes in perturbative quantum field theory. It is therefore not surprising that theoretical physicists have already used some of these structures, albeit implicitly. For example, the Berends--Giele recursion relation for currents~\cite{Berends:1987me} is nothing but a quasi-isomorphism of $L_\infty$-algebras.\footnote{Interestingly, this paper was published in the same journal in which four years later the notion of $L_\infty$-algebra was first defined~\cite{Zwiebach:1992ie}. It seems that the journal for groundbreaking research on strong homotopy Lie algebras is a journal for theoretical physics.}
    
    The bridge between Lagrangian field theories and homotopy algebras is provided by the classical part of the Batalin--Vilkovisky (BV) formalism~\cite{Batalin:1981jr,Schwarz:1992nx}, which yields a differential graded commutative algebra (dgca), usually called the {\em BV complex}, one of a number of equivalent descriptions of an $L_\infty$-algebra. Any $L_\infty$-algebra comes with a {\em homotopy Maurer--Cartan action} as well as an isomorphism class of {\em minimal models}. The homotopy Maurer--Cartan action reproduces the classical action from which the $L_\infty$-algebra was constructed, and thus any field theory is a homotopy Maurer--Cartan theory. The minimal models encode the tree level scattering amplitudes. 
    
    Since the minimal models are computed recursively, most conveniently by using the {\em homological perturbation lemma}, there are useful recursion relations underlying the computation of tree level scattering amplitudes. In the case of four-dimensional Yang--Mills theory, these are precisely the aforementioned Berends--Giele recursion relations.
    
    The full power of this homotopy algebraic perspective on field theory has not been exploited yet, and it may lead to more direct (or, indeed, first) answers to many questions in quantum field theory. It also provides a very concise explanation of the computation of scattering amplitudes in perturbative quantum field theory\footnote{We hasten to add that the homotopy algebraic perspective given here does not address regularisation and renormalisation, but we intend to formulate these issues in the context of homotopy algebras soon.} that should be more readily accessible to mathematicians. We therefore believe that it is important to explain this perspective to a broader audience and to promote its use.
    
    We begin with a short introduction to homotopy algebras. We then explain how the BV formalism assigns a homotopy algebra to any field theory and how the original action is recovered by the homotopy Maurer--Cartan theory of that homotopy algebra. Our key examples here are scalar field theory and Yang--Mills theory. We then explain how minimal models are computed with the homological perturbation lemma and how they lead to the Feynman diagrams of perturbative quantum field theory, both at tree and quantum level. We close with a short new proof for a combinatorial identity for the tree-level Berends--Giele currents in Yang--Mills theory, which implies the Kleiss--Kuijf relations~\cite{Kleiss:1988ne,DelDuca:1999rs}
    
    \subsection*{Pointers to some of the relevant original literature}
    
    The historical development of the homotopy algebraic perspective after the invention of the BV formalism becomes quickly very involved and it would be impossible to give a complete account of the literature. In the following, we point out some key references. The most important paper is certainly~\cite{Zwiebach:1992ie}, which introduced $L_\infty$-algebras, already together with their quantum variants, and established the link to the BV formalism. The closely related $A_\infty$-algebras are much older~\cite{Stasheff:1963aa,Stasheff:1963ab}. Deeper explanations of the geometry and the meaning of the classical part of the BV formalism were given by various authors in the 90ies, see e.g.~\cite{Schwarz:1992nx,Stasheff:1997fe,Stasheff:1997iz}. Concrete differential complexes underlying classical field theories were then calculated in the following decade, see e.g.~\cite{Movshev:2003ib,Movshev:2004aw,Gover:2006fa,Zeitlin:2007vv,Zeitlin:2007vd,Zeitlin:2007yf,Zeitlin:2007fp,Zeitlin:2008cc}. The fact that classical field theories have underlying $L_\infty$-algebras was later rediscovered several times, see e.g.~\cite{IuliuLazaroiu:2009wz,Hohm:2017pnh}. It also underlies much of the work of Costello, cf.~\cite{Costello:2011np}
    
    The existence of minimal models of homotopy algebras, in particular $A_\infty$- and $L_\infty$-algebras, was known for some time~\cite{kadeishvili1982algebraic}. The decomposition theorem together with a useful perturbative algorithm for the computation of the minimal model was given much later in~\cite{Kajiura:0306332}. In this reference, it was also pointed out that minimal models are related to a Feynman diagram expansion in general, following earlier suggestions going back to~\cite{Kontsevich:2000yf}. This fact was also used in~\cite{Gwilliam:2012jg} to derive Wick's theorem and Feynman rules for finite-dimensional integrals.
    
    The homological perturbation lemma~\cite{gugenheim1991perturbation,Gugenheim1989:aa,Crainic:0403266} then entered the discussion with~\cite{JohnsonFreyd:2012ww} and in particular with~\cite{Doubek:2017naz}, but its relevance seems to have been clear to several people much earlier. The extension of the homological perturbation lemma to computations of quantum minimal models is due to~\cite{Doubek:2017naz}.
    
    The explicit application to computations of S-matrices at tree level was rather recent~\cite{Arvanitakis:2019ald,Nutzi:2018vkl,Macrelli:2019afx,Reiterer:2019dys,Nutzi:2019ufl} and the generalisation to loop level is found in~\cite{Jurco:2019yfd}. 
    
    Much more detailed explanations of our formalism and its mathematical background in the conventions used here as well as a more complete list of pointers to the original literature are found in the papers~\cite{Jurco:2018sby,Jurco:2019bvp,Jurco:2019woz,Macrelli:2019afx,Jurco:2019yfd}.
    
    \section{Homotopy algebras}
    
    Homotopy algebras are generalisations of classical algebras, such as associative, Leibniz or Lie algebras, in which the relevant structural identities, that is associativity, the Leibniz identity or the Jacobi identity, respectively, hold only up to homotopies. Usually, an additional qualifier ``strong'' implies that the structural identities hold up to ``nicely behaved'' or coherent homotopies. 
    
    There are three different perspectives on homotopy algebras\footnote{A fourth one being that of algebras over operads in the category of chain complexes, which can also be useful. Particular types of strong homotopy Lie algebras are also good descriptions of categorified Lie algebras.} that are relevant to the discussion of perturbative quantum field theory: 
    \begin{enumerate}
        \item {\em higher products} or {\em higher brackets} which give rise to the formulation of an action principle;
        \item {\em codifferential graded coalgebras}, where all higher products are packaged in a single codifferential, and which is best suited for performing perturbation theory;
        \item {\em differential graded algebras}, which is the linear dual of the latter and which is the output of the classical part of the BV formalism.
    \end{enumerate}
    
    In the following, we take a brief tour through all of these descriptions and indicate how these are linked to each other.
    
    \subsection{Higher products}
    
    Consider a unital associative algebra $\fra$. A good example to have in mind is a matrix algebra which includes the identity matrix. Besides being a vector space, $\fra$ is endowed with an associative product $\sfm_2:\fra\times \fra\rightarrow \fra$ satisfying
    \begin{equation}
        \sfm_2(m_2(a_1,a_2),a_3)=\sfm_2(a_1,m_2(a_2,a_3))
    \end{equation}
    for all $a_{1,2,3}\in \fra$. A simple generalisation of this is a {\em strict $A_\infty$-algebra}, which is a differential $\IZ$-graded associative algebra,
    \begin{equation}
        \fra=\bigoplus_{i\in \IZ} \fra_i~,\hspace{1cm}\dots \xrightarrow{~\sfm_1~} \fra_{-1} \xrightarrow{~\sfm_1~} \fra_0 \xrightarrow{~\sfm_1~} \fra_{1} \xrightarrow{~\sfm_1~} \dots\hspace{1cm}\sfm_2:\fra_i\times \fra_j\rightarrow \fra_{i+j}~,
    \end{equation}
    where $\sfm_1$ is a differential which is compatible with the associative product $\sfm_2$. That is,
    \begin{equation}
        \begin{aligned}
            \sfm_1(\sfm_1(a_1))&=0~,\\
            \sfm_1(\sfm_2(a_1,a_2))&=\sfm_2(\sfm_1(a_1),a_2)+(-1)^{|a_1|}\sfm_2(a_1,\sfm_1(a_2))~,\\
            \sfm_2(m_2(a_1,a_2),a_3)&=\sfm_2(a_1,m_2(a_2,a_3))~,
        \end{aligned}
    \end{equation}
    where $|a_1|$ denotes the degree of $a_1$.
    
    To obtain a general $A_\infty$-algebra, we first have to lift associativity up to coherent homotopy by introducing a trilinear map $\sfm_3:\fra\times \fra \times \fra\rightarrow \fra$ of degree~$-1$,
    \begin{equation}
        \sfm_3:\fra_i\times \fra_j\times \fra_k\rightarrow \fra_{i+j+k-1}~,
    \end{equation}
    such that 
    \begin{equation}
        \begin{aligned}
            \sfm_2(\sfm_2(a_1,a_2),a_3)&-\sfm_2(a_1,\sfm_2(a_2,a_3))\\&=\sfm_1(\sfm_3(a_1,a_2,a_3))+\sfm_3(\sfm_1(a_1),a_2,a_3)+\\
            &\hspace{0.2cm}+(-1)^{|a_1|}\sfm_3(a_1,\sfm_1(a_2),a_3)+(-1)^{|a_1|+|a_2|}\sfm_3(a_1,a_2,\sfm_1(a_3))~.
        \end{aligned}
    \end{equation}
    More succinctly, we can write
    \begin{equation}
        \sfm_1\sfm_3-\sfm_2(\sfm_2\otimes \sfid)+\sfm_2(\sfid\otimes\sfm_2)+\sfm_3(\sfm_1\otimes \sfid\otimes \sfid+\sfid \otimes \sfm_1\otimes \sfid+\sfid\otimes \sfid\otimes \sfm_1)=0~,
    \end{equation}
    where the signs arise by passing arguments past the degree~$1$ maps $\sfm_1$. In general, we insert {\em Koszul signs}, for example:
    \begin{equation}
        (f\otimes g)(x\otimes y)=(-1)^{|g|\,|x|}(f(x)\otimes g(y))~.
    \end{equation}
    
    The complete $A_\infty$-algebra is then given by linear maps $\sfm_i:\fra^{\times i}\rightarrow \fra$ of degree~$2-i$ such that associativity holds up to coherent homotopy,
    \begin{equation}\label{eq:hJI_associative}
        \sum_{i+j+k=n}(-1)^{ij+k}\sfm_{i+k+1}(\sfid^{\otimes i}\otimes \sfm_j\otimes \sfid^{\otimes k})=0
    \end{equation}
    for all $n\in \IN^+$.
    
    The strong homotopy version of a Lie algebra is defined analogously. Here, we have a $\IZ$-graded vector space $\frg=\oplus_{i\in \IZ}\frg_i$ together with the higher products $\mu_i:\frg^{\times i}\rightarrow \frg$ which are totally antisymmetric linear maps of degree~$2-i$ such that for each $n\in \IN^+$, the {\em homotopy Jacobi identity} holds,
    \begin{equation}\label{eq:hJI_Lie}
        \begin{aligned}
            0\ &=\ \sum_{i+j=n} (-1)^{j}\mu_{j+1}\circ (\mu_i\otimes \sfid^{\otimes j})\circ \sum_{\sigma\in {\rm Sh}(i;n)} \sigma_\wedge~,
        \end{aligned}
    \end{equation}
    where the unshuffle operator $\sigma_\wedge$ creates all graded permutations of $n$ arguments $\ell_1\otimes \ell_2 \otimes \dots \otimes \ell_n$ preserving the relative order of $(\ell_{\sigma(1)},\dots,\ell_{\sigma(j)})$ and $(\ell_{\sigma(j+1)},\dots,\ell_{\sigma(n)})$ and inserting the appropriate sign for graded antisymmetry. Explicitly, the homotopy Jacobi identity reads as 
    \begin{equation}
        \sum_{i+j=n} \sum_{\sigma\in {\rm Sh}(j;i)} (-1)^{j}\chi(\sigma;\ell_1,\dots, \ell_n) \mu_{j+1}(\mu_i(\ell_{\sigma(1)},\dots,v_{\sigma(i)}),\ell_{\sigma(i+1)},\dots,\ell_{\sigma(n)})=0~,
    \end{equation}
    where $\chi(\sigma;\ell_1,\dots \ell_i)$ is the sign necessary to obtain the permutation $\sigma$ from graded antisymmetry:
    \begin{equation}\label{eq:Koszul-chi}
        (\ell_1\wedge\dots \wedge \ell_n)=\chi(\sigma;\ell_1,\ldots,\ell_n)(\ell_{\sigma(1)}\wedge \dots \wedge\ell_{\sigma(n)})~.
    \end{equation}
    
    Just as any matrix algebra comes with a natural Lie bracket given by the commutator, any $A_\infty$-algebra comes with the higher products of an $L_\infty$-algebra,
    \begin{equation}\label{eq:A_infty_to_L_infty}
        \mu_i(\ell_1,\dots,\ell_n)=\left(\sfm_i\circ \sum_{\sigma \in S_n} \sigma_\wedge\right)(\ell_1,\dots,\ell_n)=\sum_{\sigma \in S_n}\chi(\sigma;\ell_1,\dots \ell_i)\sfm_i(\ell_{\sigma(1)},\dots ,\ell_{\sigma(n)})~,
    \end{equation}
    which is reasonably obvious comparing~\eqref{eq:hJI_associative} and~\eqref{eq:hJI_Lie}. If the $A_\infty$-algebra is just a matrix algebra, the corresponding $L_\infty$-algebra is indeed just the usual matrix Lie algebra induced by the commutator.
    
    We note that the definition of both $A_\infty$- and $L_\infty$-algebras readily extends from graded vector spaces to graded modules over rings over $\IR$.  
    
    \subsection{Codifferential graded coalgebras and differential graded algebras}
    
    The higher products $\sfm_i$ and $\mu_i$ introduced above for $A_\infty$- and $L_\infty$-algebras are always of degree~$2-i$. If we shift the underlying graded vector spaces $\fra$ and $\frg$ by~$-1$, switching to $\fra[1]$ and $\frg[1]$, then the degree of all higher products changes to 1. We use the usual notation,
    \begin{equation}
        \fra[k]=\bigoplus_{i\in\IZ} (\fra[k])_i~~~\mbox{with}~~~\fra[k]_i=\fra_{k+i}~.
    \end{equation}
    Once this is done, we can continue the higher products $\sfm_i$ and $\mu_i$ to coderivations on the tensor algebras\footnote{Strictly speaking, one either needs to remove the summand $\IR$ from the tensor algebras or admit {\em curved} homotopy algebras, containing a higher product $\sfm_0$ or $\mu_0$. For simplicity, we ignore this point in our discussion.}
    \begin{equation}
        \begin{aligned}
            \bigotimes{}^\bullet\fra[1]&=\IR~\oplus~\fra[1]~\oplus~\fra[1]\otimes\fra[1]~\oplus~\fra[1]\otimes\fra[1]\otimes\fra[1]~\oplus~\dots~,\\
            \bigodot{}^\bullet\frg[1]&=\IR~\oplus~\frg[1]~\oplus~\frg[1]\odot\frg[1]~\oplus~\frg[1]\odot\frg[1]\odot\frg[1]~\oplus~\dots~,
        \end{aligned}
    \end{equation}
    respectively. Recall that a {\em coderivation} $D$ satisfies the {\em co-Leibniz rule}
    \begin{equation}\label{eq:coLeibniz}
        \Delta\circ D=(D\otimes \sfid)\circ \Delta+(\sfid\otimes D)\circ \Delta~,
    \end{equation}
    where the relevant coproducts $\Delta_\otimes$ and $\Delta_\odot$ on $\bigotimes{}^\bullet\fra[1]$ and $\bigodot{}^\bullet\frg[1]$ are the shuffle coproduct and its symmetrised form:
    \begin{equation}
        \begin{aligned}
            \Delta_\otimes(a_1\otimes \dots \otimes a_n)&=\sum_{k=0}^n (a_1\otimes \dots \otimes a_k)\bigotimes(a_{k+1}\otimes \dots \otimes a_k)~,\\
            \Delta_\odot(\ell_1\odot \dots \odot \ell_n)&=\sum_{k=0}^n\sum_{\sigma\in {\rm Sh}(k;n)}\eps(\sigma;\ell_1,\ldots,\ell_n)(\ell_{\sigma(1)}\odot \dots \odot \ell_{\sigma(k)})\bigotimes (\ell_{\sigma(j+1)}\odot \dots \odot \ell_{\sigma(n)})
        \end{aligned}
    \end{equation}
    for $a_i\in \fra[1]$ and $\ell_i\in \frg[1]$, where $\eps(\sigma;\ell_1,\dots \ell_i)$ is the sign necessary to obtain the permutation $\sigma$ from graded symmetry:
    \begin{equation}
        (\ell_1\odot \dots \odot \ell_n)=\eps(\sigma;\ell_1,\ldots,\ell_n)(\ell_{\sigma(1)}\odot \dots \odot \ell_{\sigma(n)})~.
    \end{equation}
    
    For an $A_\infty$-algebra, for example, the continuation of $\sfm_2$ to a coderivation $\sfm_2$ on $\bigotimes{}^\bullet\fra[1]$ reads as
    \begin{equation}
        \begin{aligned}
            &\sfm_2(r)=0~,~~~\sfm_2(a_1)=0~,~~~\sfm_2(a_1\otimes a_2)=\sfm_2(a_1,a_2)~,\\
            &\sfm_2(a_1\otimes a_2\otimes a_3)=\sfm_2(a_1,a_2)\otimes a_3+(-1)^{|a_1|}a_1\otimes \sfm_2(a_2,a_3)~,~~~\dots
        \end{aligned}
    \end{equation}
    for all $r\in \IR$, $a_{1,2,3}\in \fra[1]$. Note that we slightly abused notation, using the same symbol for $\sfm_2$ and the map it induces on $\fra[1]$. Also, Koszul signs arise as above. We now combine the higher products into single coderivations of degree~$1$:
    \begin{equation}
        D_\otimes=\sfm_1+\sfm_2+\sfm_3+\dots\eand            D_\odot=\mu_1+\mu_2+\mu_3+\dots~.
    \end{equation}
    In both cases, the identity for associativity up to coherent homotopy~\eqref{eq:hJI_associative} as well as the homotopy Jacobi identity~\eqref{eq:hJI_Lie} amount to $D_\otimes$ and $D_\odot$ being differentials:
    \begin{equation}
        D_\otimes^2=0\eand D_\odot^2=0~.
    \end{equation}
    We thus saw that $A_\infty$-algebras correspond to codifferential graded (cofree\footnote{which implies that they arise as the tensor algebra on some graded vector space}) coalgebras or codgcoas for short. In the case of $L_\infty$-algebras, the codgcoas are also cocommutative.
    
    In many cases, in particular when the homogeneously graded subspaces of $\frg$ and $\fra$ are finite dimensional, one can dualise this picture. The result is a differential graded algebra or dga for short. Explicitly, we consider the algebras
    \begin{equation}
        \begin{aligned}
            \bigotimes{}^\bullet\fra[1]^*&=\IR~\oplus~\fra[1]^*~\oplus~\fra[1]^*\otimes\fra[1]^*~\oplus~\fra[1]^*\otimes\fra[1]^*\otimes\fra[1]^*~\oplus~\dots~,\\
            \bigodot{}^\bullet\frg[1]^*&=\IR~\oplus~\frg[1]^*~\oplus~\frg[1]^*\odot\frg[1]^*~\oplus~\frg[1]^*\odot\frg[1]^*\odot\frg[1]^*~\oplus~\dots~,
        \end{aligned}
    \end{equation}
    and the codifferentials $D_\odot$ and $D_\otimes$ dualise to differentials 
    \begin{equation}
        Q_\odot \coloneqq D_\odot^*\eand Q_\otimes \coloneqq D_\otimes^*
    \end{equation}
    with 
    \begin{equation}
        Q_\odot^2=0\eand Q_\otimes^2=0~.
    \end{equation}
    To physicists, these differentials are familiar, e.g.~from the BRST and BV complexes.
    
    \subsection{Examples}
    
    To illustrate the above constructions, let us briefly consider the example of an ordinary matrix algebra $\fra=\fra_0$. The codifferential $D_\otimes$ on $\bigotimes^\bullet\fra[1]$ is simply given by the matrix product, continued to a coderivation. For example,
    \begin{equation}
        D_\otimes(a_1\otimes a_2\otimes a_3\otimes a_4) \coloneqq a_1a_2\otimes a_3\otimes a_4-a_1\otimes a_2a_3\otimes a_4+a_1\otimes a_2\otimes a_3a_4~,
    \end{equation}
    where $a_i\in \fra[1]$. Clearly, $D_\otimes^2=0$ amounts to associativity, for example
    \begin{equation}
        \begin{aligned}
            D_\otimes(D_\otimes(a_1\otimes a_2\otimes a_3))& \coloneqq D_\otimes(a_1a_2\otimes a_3-a_1\otimes a_2a_3)\\
            &\phantom{:}=(a_1a_2)a_3-a_1(a_2a_3)~.
        \end{aligned}
    \end{equation}
    On $\bigodot^\bullet\frg[1]$ for $\frg=\fra$, the symmetrisation in shifted degrees leads to the coproduct 
    \begin{equation}
        D_\odot(a_1\odot a_2\odot a_3)\coloneqq (a_1a_2-a_2a_1)\odot a_3-(a_1a_3-a_3a_1)\odot a_2+(a_2a_3-a_3a_2)\odot a_1~,
    \end{equation}
    where we used the fact that the $a_i\in \fra[1]$ are all of degree~$-1$. Here, $D_\odot^2=0$ leads to the Jacobi identity:
    \begin{equation}
        \begin{aligned}
            D_\odot(D_\odot(a_1\odot a_2\odot a_3))& \coloneqq D_\odot([a_1,a_2]\odot a_3-[a_1,a_3]\odot a_2+[a_2,a_3]\odot a_1)\\
            &\phantom{:}=[[a_1,a_2],a_3]-[[a_1,a_3],a_2]+[[a_2,a_3],a_1]~.
        \end{aligned}
    \end{equation}
    
    To simplify the dualisation, let us assume that $\fra$ is finite-dimensional and introduce a basis $\tau_a$, $a=1,\dots,\dim(\fra)$ on $\fra$. We then have structure constants
    \begin{equation}
        \tau_a\tau_b=:s^c_{ab}\tau_c\eand [\tau_a,\tau_b]=:f^c_{ab}\tau_c~.
    \end{equation}
    Together with the coordinate functions $\xi^a:\fra[1]\rightarrow \IR$ dual to the shifted basis $\tau_a$ on $\fra[1]$, we have the differentials
    \begin{equation}
        Q_\otimes \xi^a=-s^a_{bc}\xi^b\otimes \xi^c\eand Q_\odot \xi^a=-\tfrac12 f^a_{bc} \xi^b\odot \xi^c~,
    \end{equation}
    which both satisfy the Leibniz rule,
    \begin{equation}
        \begin{aligned}
            Q_\otimes (p_1\otimes p_2)&=(Q_\otimes p_1)\otimes p_2+(-1)^{|p_1|}p_1\otimes (Q_\otimes p_2)~,\\
            Q_\odot (p_1\odot p_2)&=(Q_\odot p_1)\odot p_2+(-1)^{|p_1|}p_1\odot (Q_\odot p_2)
        \end{aligned}
    \end{equation}
    for any elements $p_{1,2}$ of $\bigotimes^\bullet\fra[1]$ or $\bigodot^\bullet\fra[1]$, respectively. The fact that $Q_\otimes$ and $Q_\odot$ are differentials, and thus square to zero, corresponds again to associativity and the Jabobi identity for the commutator, respectively.
    
    Genuine $A_\infty$- and $L_\infty$-algebras arise in a number of contexts in physics. We postpone discussing any further examples to section~\ref{sec:homotopy_algebras_and_field_theories}, where homotopy algebras arise in the form of BRST and BV complexes of classical field theories.
    
    \subsection{Cyclic structures}
    
    To write down action principles for usual gauge theories requires an invariant metric on the gauge Lie algebra, i.e.~the data of a quadratic Lie algebra. There is a straightforward generalisation to a differential graded Lie algebra $\frg$, where an inner product is a bilinear, graded symmetric, and non-degenerate map $\langle -,-\rangle:\frg\times \frg\rightarrow \IR$ satisfying
    \begin{equation}\label{eq:cyclicitydgL}
        \begin{aligned}
            \langle \rmd \ell_1,\ell_2\rangle+(-1)^{|\ell_1|}\langle \ell_1,\rmd \ell_2\rangle&=0~,\\
            \langle [\ell_1,\ell_2],\ell_3\rangle+(-1)^{|\ell_1|\,|\ell_2|}\langle \ell_2,[\ell_1,\ell_3]\rangle&=0~.
        \end{aligned}
    \end{equation}
    These relations have a straightforward generalisation to $L_\infty$-algebras:
    \begin{equation}\label{eq:cyclicity_L}
        \langle \mu_i(\ell_1,\dots,\ell_{i-1},\ell_i),\ell_{i+1}\rangle+(-1)^{|\ell_{i}|(i+|\ell_1|+\dots+|\ell_{i-1}|)}\langle \ell_{i},\mu_i(\ell_1,\dots,\ell_{i-1},\ell_{i+1})\rangle=0~,
    \end{equation}
    which is equivalent to
    \begin{equation}\label{eq:cyclicity_L2}
        \langle\ell_1,\mu_i(\ell_2,\ldots,\ell_{i+1})\rangle=(-1)^{i+i(|\ell_1|+|\ell_{i+1}|)+|\ell_{i+1}|\sum_{j=1}^{i}|\ell_j|}\langle\ell_{i+1},\mu_i(\ell_1,\ldots,\ell_{i})\rangle
    \end{equation}
    for all $\ell_i\in \frg$. Such inner products are called {\em cyclic structures} and an $L_\infty$-algebra endowed with a cyclic structure is simply called {\em cyclic}.
    
    This definition does not rely on symmetrisation, and it thus also applies to $A_\infty$-algebras. That is, a {\em cyclic $A_\infty$-algebra} is an $A_\infty$-algebra $(\fra,\sfm_i)$ with a bilinear, graded symmetric, and non-degenerate map $\langle -,-\rangle:\fra\times \fra\rightarrow \IR$ satisfying 
    \begin{equation}\label{eq:cyclicity_A}
        \langle a_1,\sfm_i(a_2,\ldots,a_{i+1})\rangle=(-1)^{i+i(|a_1|+|a_{i+1}|)+|a_{i+1}|\sum_{j=1}^{i}|a_j|}\langle a_{i+1},\sfm_i(a_1,\ldots,a_{i})\rangle
    \end{equation}
    for all $a_i\in \fra$. Note that a strict $A_\infty$-algebras is nothing but a {\em differential (noncommutative) Frobenius algebra}.

    \section{Homotopy algebras from classical field theories}\label{sec:homotopy_algebras_and_field_theories}
    
    In this section, we show how the action of a gauge group and the equations of motion {\em both} lead to differentials that nicely combine into the BV differential. The resulting BV complex is dual to an $L_\infty$-algebra, and thus field theories correspond to strong homotopy Lie algebras. This assignment is ``functorial'' in the sense that embeddings, projections and equivalences of field theories map to corresponding morphisms of $L_\infty$-algebras. Moreover, any $L_\infty$-algebra comes with a natural ``field theory,'' called {\em homotopy Maurer--Cartan theory}, and the homotopy Maurer--Cartan action of the $L_\infty$-algebra of a field theory is just the field theory's original action.
    
    \subsection{Homotopy Maurer--Cartan theory}
    
    Recall that a Maurer--Cartan element $a$ of a differential graded Lie algebra $(\frg,\rmd,[-,-])$ is an element of degree~$1$ satisfying the Maurer--Cartan equation
    \begin{equation}
        \rmd a+\tfrac12 [a,a]=0~.
    \end{equation}
    If $\frg$ is endowed with an invariant inner product $\langle -,-\rangle$ compatible with the differential, this equation is the equation of motion of the Maurer--Cartan action
    \begin{equation}
        S_{\rm MC}=\tfrac12\langle a,\rmd a\rangle+\tfrac{1}{3!}\langle a,[a,a]\rangle~,
    \end{equation}
    where $\langle-,-\rangle$ is an inner product on $\frg$ satisfying~\eqref{eq:cyclicitydgL}.
    
    The evident generalisation of this action to {\em homotopy Maurer--Cartan theory} reads as
    \begin{equation}\label{eq:hMCAction}
        S_{\rm hMC}[a] \coloneqq \sum_{i\geq1} \frac{1}{(i+1)!}\langle a,\mu_i(a,\ldots,a)\rangle~,
    \end{equation}
    where we call $a\in \frg_1$ the {\em gauge potential}. Its {\em curvature} is defined as 
    \begin{equation}\label{eq:Curvature}
        f \coloneqq \mu_1(a)+\tfrac12 \mu_2(a,a)+\cdots=\sum_{i\geq 1}\frac{1}{i!}\mu_i(a,\ldots,a)~,
    \end{equation}
    and $f=0$ are the stationary points of~\eqref{eq:hMCAction}, defining homotopy Maurer--Cartan elements $a$. The homotopy Jacobi identities induce a Bianchi identity,
    \begin{equation}\label{eq:BianchiIdentity}
        \sum_{i\geq0}\frac{(-1)^i}{i!}\mu_{i+1}(f,a,\ldots,a)=0~,
    \end{equation}
    and the action~\eqref{eq:hMCAction} is invariant under the infinitesimal gauge transformations
    \begin{equation}\label{eq:GaugeTrafo}
        \delta_{ c_0} a=\sum_{i\geq0} \frac{(-1)^i}{i!}\mu_{i+1}(c_0,a,\ldots,a)~,
    \end{equation}
    where $c_0\in \frg_0$ is the gauge parameter. There are also higher gauge transformations; for details see~\cite{Jurco:2018sby}. Similarly, there are higher Bianchi identities.
    
    We conclude that any $L_\infty$-algebra $\frg$ comes with the full kinematical data of a gauge theory, together with equations of motion $f=0$. If $\frg$ is cyclic, the equations of motion arise from an action principle. Clearly, the theory is essentially trivial, if there are no gauge potentials, which is the case for $\frg_1=*$. We note the following roles of the homogeneously graded subspaces $\frg_i$ of $\frg$:
    \begin{itemize}
        \item[$\circ$] the gauge potential or the general field takes values in $\frg_1$;
        \item[$\circ$] the curvature takes values in $\frg_2$;
        \item[$\circ$] the ``left-hand side'' of the Bianchi identity takes values in $\frg_3$;
        \item[$\circ$] gauge symmetries are parametrised by $\frg_0$;
        \item[$\circ$] higher gauge symmetries are parametrised by $\frg_i$, $i<0$, and the ``left-hand sides'' of higher Bianchi identities take values in $\frg_i$, $i>3$.
    \end{itemize}
    
    A homotopy Maurer--Cartan action also exists for $A_\infty$-algebras $\fra$, where we similarly have a gauge potential $a\in \fra_1$ with curvature
    \begin{equation}\label{eq:Curvature_A}
        f \coloneqq \sfm_1(a)+\sfm_2(a,a)+\cdots=\sum_{i\geq 1}\sfm_i(a,\ldots,a)~,
    \end{equation}
    satisfying the Bianchi identity
    \begin{equation}\label{eq:BianchiIdentity_A}
        \sum_{i\geq 0}\sum_{j=0}^{i-1}(-1)^{i+j}\sfm_{i+1}(\underbrace{a,\ldots,a}_{j},f,\underbrace{a,\ldots,a}_{i-j})=0~.
    \end{equation}
    Flatness $f=0$ arises as the equation of motion of the action
    \begin{equation}\label{eq:hMCAction_A}
        S_{\rm hMC}[a] \coloneqq \sum_{i\geq1} \frac{1}{i+1}\langle a,\sfm_i(a,\ldots,a)\rangle~,
    \end{equation}
    which is invariant under infinitesimal gauge transformations
    \begin{equation}\label{eq:GaugeTrafo_A}
        \delta_{ c_0} a=\sum_{i\geq0}\sum_{j=0}^{i-1}(-1)^{i+j} \sfm_{i+1}(\underbrace{a,\ldots,a}_{j}, c_0,\underbrace{a,\ldots,a}_{i-j})~,
    \end{equation}
    where $c_0\in\fra_0$ again parametrises the gauge transformations. Note that these formulas yield those for an $L_\infty$-algebra after graded antisymmetrisation~\eqref{eq:A_infty_to_L_infty}.
    
    \subsection{The BV formalism}
    
    The BV formalism is commonly used in the quantisation of field theories with ``open'' gauge symmetries, i.e.~gauge symmetries that only close on-shell. This is the case e.g.~in ordinary higher gauge theories. Usually, constructing the BV complex is a two-step process. In a first step, one introduces {\em ghosts}, which are fermionic fields parametrising infinitesimal gauge transformations. This leads to the BRST differential and the BRST complex, which encodes the action of gauge transformations (as well as higher gauge transformations encoded in ghosts for ghosts) on the various fields. The BV complex is then obtained by introducing an {\em antifield} for each field, ghost, ghost for ghosts, etc., together with a graded Poisson bracket $\{-,-\}_{\rm BV}$ called {\em antibracket}. One then extends the classical action to the (classical) BV action $S_{\rm BV}=S_{\rm classical}+\dots$ by terms containing the antifields such that {\em classical master equation}
    \begin{equation}
        \{S_{\rm BV},S_{\rm BV}\}=0
    \end{equation}
    holds. This recipe appears quite ad-hoc, but regarded from a more mathematical perspective, the situation becomes much clearer. The aim of the classical part of the BV formalism is to gain a good description of classical observables. These are functions on field space modulo gauge transformations and restricted to fields satisfying the equations of motion. 
    
    The first step is to quotient by gauge symmetries. We know that quotient spaces can be problematic to deal with, so it is better to consider ``derived quotients''. That is, we encode the whole gauge structure in an action groupoid, i.e.~a category with invertible morphisms, where the objects are the field configurations and the morphisms are gauge transformations. Differentiating an action groupoid yields an action Lie algebroid (a special kind of $L_\infty$-algebroid), and if the field space is regarded as a vector space, we obtain an $L_\infty$-algebra. In the special case of an ordinary gauge theory, this $L_\infty$-algebra is of the form $\frg_{\rm BRST}=\frg_0\oplus \frg_1$, where $\frg_1$ are the fields and $\frg_0$ the infinitesimal gauge transformations. The dual description of this $L_\infty$-algebra in terms of differential graded algebras is nothing but the usual BRST complex, and this is known as a Chevalley--Eilenberg resolution of the functions on field space modulo gauge transformations. Here, gauge-trivial observables are functions on field space which are $Q_{\rm BRST}$-exact and gauge invariant observables are functions which are $Q_{\rm BRST}$-closed.
    
    To restrict to fields satisfying the equations of motion, we perform a Koszul--Tate resolution. That is, we enlarge the $L_\infty$-algebra $\frg_{\rm BRST}$ to its cotangent space $\frg_{\rm BV}=T^*[-1]\frg_{\rm BRST}$ with the duals of the fields, ghosts and higher ghosts called antifields (and antifields of ghosts, etc.). As a cotangent space, $\frg_{\rm BV}$ comes with the natural symplectic form 
    \begin{equation}
        \omega_{\rm BV}=\rmd \Phi^A\wedge \rmd\Phi^+_A~,
    \end{equation}
    where $\Phi^A$ is a coordinate function on $\frg_{\rm BRST}$. This implies that $A$ is a multi-index running over fields, ghosts, and their antighosts  as well as all their labels such as momenta, tensor and gauge labels. This symplectic form induces the {\em antibracket}, a Poisson bracket $\{-,-\}_{\rm BV}$ of degree~$1$.\footnote{Poisson algebras with Poisson brackets of degree~$1$ are also known as {\em Gerstenhaber algebras}.} Moreover, we extend $Q_{\rm BRST}$ to a Hamiltonian vector field $Q_{\rm BV}=\{S_{\rm BV},-\}_{\rm BV}$, where $S_{\rm BV}$ is the minimal extension of the classical action $S_{\rm cl}$ such that $Q_{\rm BV}$ restricts to $Q_{\rm BRST}$ for $\Phi^+_A=0$ and such that $S_{\rm BV}$ satisfies the {\em classical master equation}
    \begin{equation}
        \{S_{\rm BV},S_{\rm BV}\}=0~.
    \end{equation}
    It follows that $Q_{\rm BV}$ encodes the equation of motion obtained by varying the classical action with respect to the field $\phi$:
    \begin{equation}
        Q_{\rm BV} \phi^+=\{S_{\rm BV}, \phi^+\}=\frac{\delta S_{\rm BV}}{\delta \phi}~.
    \end{equation}
    In the absence of ghosts, the right hand side is the equation of motion and functions on field space differing by $Q_{\rm BV}$-exact terms are on-shell equivalent. If ghosts are added to the picture, then such functions are on-shell equivalent after taking into account gauge symmetries. Observables are thus encoded in the $Q_{\rm BV}$-cohomology.
    
    Altogether, the BV formalism assigns to any Lagrangian field theory an $L_\infty$-algebra given by the dga encoded in the BV complex. In practice, however, it will be much more convenient to derive the $L_\infty$-algebra directly: either by a direct construction, or by matching the action and gauge transformations to a homotopy Maurer--Cartan theory. We shall discuss this in the following.
    
    \subsection{Higher Chern--Simons theories}
    
    Homotopy Maurer--Cartan theory is clearly a vast generalisation of Chern--Simons theory, and one can derive higher Chern--Simons theories directly by constructing suitable $L_\infty$-algebras, cf.~\cite{Saemann:2017vuy}.
    
    We first observe that the tensor product of a differential graded algebra and an $L_\infty$-algebra comes with a natural $L_\infty$-algebra structure. For concreteness' sake, let us focus on the case where the differential graded algebra is the de~Rham complex $(\Omega^\bullet(M),\rmd)$ on some manifold $M$. Then the $L_\infty$-algebra $\Omega^\bullet(M,\frg)=\Omega^\bullet(M)\otimes \frg$ has higher products
    \begin{equation}
        \begin{aligned}
            \hat \mu_1 &\coloneqq \rmd \otimes \sfid+\sfid\otimes \mu_1~,\\
            \hat \mu_i &\coloneqq \sfid\otimes \mu_i~~~\mbox{for}~~i\geq 2~.
        \end{aligned}
    \end{equation}
    Note that the total degree of elements in $\Omega^\bullet(M,\frg)$ is the sum of the individual degrees.
    
    If $M$ is compact and $\frg$ is a cyclic $L_\infty$-algebra, then $\Omega^\bullet(M,\frg)$ is also cyclic with
    \begin{equation}
        \langle -,-\rangle_{\Omega^\bullet(M,\frg)} \coloneqq \int_M \langle-,-\rangle_\frg~.
    \end{equation}
    
    The homotopy Maurer--Cartan action for $\Omega^\bullet(M,\frg)$ now provides a higher Chern--Simons theory on $M$. For the gauge potentials to be connections on higher non-abelian gerbes, one should restrict $\frg$ to be trivial in positive degrees: $\frg_i=*$ for $i>0$. This amounts to restricting $\frg$ to a categorified Lie algebra. Moreover, if the manifold $M$ is of dimension $n+3$, then $\frg$ should be non-trivial in degrees $k$ for $-n\leq k\leq 0$.
    
    As an example, consider a four-dimensional manifold $M$ and an $L_\infty$-algebra of the form $\frg=\frg_{-1}\oplus \frg_0$. The gauge potential decomposes as
    \begin{equation}
        a=A+B\in \Omega^1(M,\frg_0)\oplus \Omega^2(M,\frg_{-1})~,
    \end{equation}
    and its curvature reads as
    \begin{equation}
        \begin{aligned}
            f&=\hat\mu_1(a)+\tfrac12\hat\mu_2(a,a)+\tfrac1{3!}\hat\mu_3(a,a,a)=F+H\in \Omega^2(M,\frg_0)\oplus \Omega^3(M,\frg_{-1})~,\\
            F&=\rmd A+\tfrac12 \mu_2(A,A)+\mu_1(B)~,\\
            H&=\rmd B+\mu_2(A,B)-\tfrac{1}{3!}\mu_3(A,A,A)~,
        \end{aligned}
    \end{equation}
    where the higher products $\mu_3$ here do not ``see'' the form part of a gauge potential. The homotopy Maurer--Cartan action is then higher Chern--Simons theory\footnote{This is the theory for a topologically trivial underlying non-abelian gerbe. Just as in the case of ordinary Chern--Simons theory, one can generalise this action and glue together this local description to a global picture.} on a four-dimensional manifold $M$,
    \begin{equation}\label{eq:4dHCSAction}
        S_{\rm hCS}=\int_M\Big\{\langle B,\rmd A+\tfrac12\mu_2(A,A)+\tfrac12\mu_1(B)\rangle+\tfrac{1}{4!}\langle\mu_3(A,A,A), A\rangle\Big\}~.
    \end{equation}
    
    Our construction here can be regarded as a concrete example of the AKSZ-formalism~\cite{Alexandrov:1995kv}, see~\cite{Jurco:2018sby} for a brief summary. This formalism directly yields a BV complex from two differential graded algebras representing a source and target space.
    
    \subsection{Examples}
    
    It turns out that we can usually circumvent the BV procedure and obtain rather directly an $A_\infty$-algebra, by matching the action of the field theory under consideration to the homotopy Maurer--Cartan action. Also, it is often more convenient to work with $A_\infty$-algebras which become the $L_\infty$-algebra that would be obtained from the BV formalism after graded antisymmetrisation~\eqref{eq:A_infty_to_L_infty}.
    
    As a first example, consider scalar field theory with cubic and quartic interactions on four-dimensional Minkowski space $\IR^{1,3}$ with metric $\eta$,
    \begin{equation}\label{eq:phi4-action}
        S_{\rm scalar} \coloneqq -\int\rmd^{4}x~\Big\{\tfrac12\varphi\Box\varphi+\tfrac{\kappa}{3!}\varphi^3+\tfrac{\lambda}{4!}\varphi^4\Big\}~,
    \end{equation}
    where $\Box \coloneqq \eta^{\mu\nu}\partial_\mu\partial_\nu$ and $\kappa,\lambda\in\IR$. The graded vector space underlying the relevant $A_\infty$-algebra is~\cite{Jurco:2019yfd}
    \begin{subequations}\label{eq:a_infty_scalar}
        \begin{equation}
            \fra=\fra_1\oplus \fra_2 \coloneqq C^\infty(\IR^{1,3})\oplus C^\infty(\IR^{1,3})
        \end{equation}
        and the non-trivial higher products are maps $\sfm_i:\fra_1^{\times i}\rightarrow \fra_2$,
        \begin{equation}\label{eq:higher_products}
            \begin{gathered}
                \sfm_1(\varphi_1) \coloneqq -\Box\varphi_1~,~~
                \sfm_2(\varphi_1,\varphi_2) \coloneqq -\kappa\varphi_1\varphi_2~,~~~\sfm_3(\varphi_1,\varphi_2,\varphi_3) \coloneqq -\lambda \varphi_1\varphi_2\varphi_3~,
            \end{gathered}
        \end{equation}
        for $\varphi_i\in\fra_1$. The cyclic structure is 
        \begin{equation}
            \langle \varphi,\varphi^+\rangle \coloneqq \int\rmd^{4}x~\varphi(x)\varphi^+(x)
        \end{equation}
    \end{subequations}
    for $\varphi\in \fra_1$ and $\varphi^+\in \fra_2$. One readily checks that the homotopy Maurer--Cartan action~\eqref{eq:hMCAction_A} for this $A_\infty$-algebra reproduces~\eqref{eq:phi4-action}.
    
    A more interesting example is Yang--Mills theory on $\IR^{1,3}$. We start from the usual action
    \begin{equation}\label{eq:YM_action_0}
        S_{\rm YM} \coloneqq \int~{\rm tr}\,\Big\{\tfrac12 F\wedge\star F\Big\}\ewith F=\rmd A+\tfrac12[A,A]
    \end{equation}
    for a gauge potential one form $A$ taking values in a gauge matrix Lie algebra $\frh$. Here, we know that we should also take into account gauge transformations, and the graded vector space underlying the $A_\infty$-algebra is thus of the form
    \begin{subequations}\label{eq:a_infty_YM}
        \begin{equation}
            \fra=\fra_0\oplus \fra_1\oplus \fra_2\oplus \fra_3 \coloneqq \Omega^0(M,\frg)\oplus\Omega^1(M,\frg)\oplus\Omega^1(M,\frg)\oplus\Omega^0(M,\frg)~,
        \end{equation}
        where we will have ghosts $c_i\in \fra_0$, fields $A_i\in \fra_1$, antifields $A_i^+\in \fra_2$ and antifields to the ghosts $c_i^+\in \fra_3$. We define the evident inner product
        \begin{equation}\label{eq:YM_cyclic_1}
            \begin{aligned}
                &\langle c_1+A_1+A_1^++c_1^+,c_2+A_2+A_2^++c_2^+\rangle\\
                &\hspace{3cm} \coloneqq \int_{\IR^{1,3}}{\rm tr}\,\Big\{c^\dagger_1\wedge\star c^+_2+c_2^\dagger\wedge\star c^+_1
                -A^\dagger_1\wedge\star A^+_2-A^\dagger_2\wedge\star A^+_1
                \Big\}
            \end{aligned}
        \end{equation}
        on $\fra$. The differential $\sfm_1$ now encodes linearised gauge transformations and equations of motions, completed cyclically:
        \begin{equation}
            \sfm_1(c_1)=-\rmd c_1\in \fra_1~,~~~
            \sfm_1(A_1)=-\rmd^\dagger \rmd A_1\in \fra_2~,~~~
            \sfm_1(A^+_1)=-\rmd^\dagger A^+_1\in \fra_3~.
        \end{equation}
        The non-abelian terms are incorporated by the higher products,
        \begin{equation}
            \begin{aligned}
                \sfm_2(c_1+A_1+A^+_1+&c^+_1,c_2+A_2+A_2^++c^+_2)&\\
                &=\underbrace{\kappa c_1c_2}_{\in \fra_0}+\underbrace{\kappa (c_1A_2+A_1c_2)}_{\in \fra_1}+\underbrace{\kappa (-c_1A_2^++A_1^+c_2)}_{\in \fra_2}+\\
                &~~~~+\underbrace{\kappa (\rmd^\dagger(A_1\wedge A_2)+\star (A_1\wedge \star\rmd A_2)-\star(\star(\rmd A_1\wedge A_2))}_{\in \fra_2}\\
                &~~~~+\underbrace{\kappa (c_1c_2^+-c_2^+c_1)+\kappa (-\star(A_1\wedge \star A_2^+)+\star(A_1^+\wedge \star A_2))}_{\in \fra_3}~,\\
                \sfm_3(A_1,A_2,A_3)&=\kappa^2(\star(A_1\wedge\star(A_2\wedge A_3))}-{\star(\star(A_1\wedge A_2)\wedge A_3)\in \fra_2~.
            \end{aligned}
        \end{equation}
    \end{subequations}
    These follow from non-abelian gauge transformations, field equations and completion via the cyclic structure~\eqref{eq:YM_cyclic_1}. Again, the homotopy Maurer--Cartan action for $\fra$ reproduces the Yang--Mills action~\eqref{eq:YM_action_0}.
    
    \section{Equivalence of field theories}
    
    The correspondence between Lagrangian field theories and $L_\infty$-algebras is functorial, and in particular, physical equivalence of classical field theories amounts to equivalence between $L_\infty$-algebras. The appropriate mathematical notion for this is a {\em quasi-isomorphism}.
    
    \subsection{Quasi-isomorphisms}
    
    Both $A_\infty$- and $L_\infty$-algebras are differential complexes, and one may be tempted to define morphisms between two such homotopy algebras as chain maps between the underlying complexes, which respect the higher products. These morphisms are called {\em strict}, but they are too restrictive for most purposes.
    
    A more appropriate notion is found in the differential graded algebra and codifferential graded coalgebra descriptions, where there is an obvious notion of morphism. This notion contains the above mentioned strict morphisms. Translated back to the homotopy algebras, morphisms
    \begin{equation}
        \fra \xrightarrow{~\phi~}\tilde \fra \eand \frg \xrightarrow{~\psi~}\tilde \frg
    \end{equation}
    are encoded in maps 
    \begin{equation}
        \phi=(\phi_i:\fra^{\times i}\rightarrow \fra)~,~~~|\phi_i|=1-i\eand \psi=(\psi_i:\frg^{\wedge i}\rightarrow \frg)~,~~~|\psi_i|=1-i
    \end{equation}
    for $i\in \IN^+$. These link the higher products in the two homotopy algebras. For example, in the case of $L_\infty$-algebras, we have the following relation:
    \begin{subequations}\label{eq:L_infty_morphism}
        \begin{equation}
            \begin{aligned}
                &\sum_{j+k=i}\sum_{\sigma\in {\rm Sh}(j;i)}~(-1)^{k}\chi(\sigma;\ell_1,\ldots,\ell_i)\psi_{k+1}(\mu_j(\ell_{\sigma(1)},\dots,\ell_{\sigma(j)}),\ell_{\sigma(j+1)},\dots ,\ell_{\sigma(i)})\\
                \ &=\ \sum_{j=1}^i\frac{1}{j!} \sum_{k_1+\cdots+k_j=i}\sum_{\sigma\in{\rm Sh}(k_1,\ldots,k_{j-1};i)}\chi(\sigma;\ell_1,\ldots,\ell_i)\zeta(\sigma;\ell_1,\ldots,\ell_i)\,\times\\
                &\kern1cm\times \tilde \mu_j\Big(\psi_{k_1}\big(\ell_{\sigma(1)},\ldots,\ell_{\sigma(k_1)}\big),\ldots,\psi_{k_j}\big(\ell_{\sigma(k_1+\cdots+k_{j-1}+1)},\ldots,\ell_{\sigma(i)}\big)\Big)~,
            \end{aligned}
        \end{equation}
        where $\chi(\sigma;\ell_1,\ldots,\ell_i)$ is again the Koszul sign~\eqref{eq:Koszul-chi} and
        \begin{equation}
            \zeta(\sigma;\ell_1,\ldots,\ell_i) \coloneqq (-1)^{\sum_{1\leq m<n\leq j}k_mk_n+\sum_{m=1}^{j-1}k_m(j-m)+\sum_{m=2}^j(1-k_m)\sum_{k=1}^{k_1+\cdots+k_{m-1}}|\ell_{\sigma(k)}|}~.
        \end{equation}
    \end{subequations}
    
    We then have two notions of isomorphism. First, an {\em isomorphism} of $A_\infty$- or $L_\infty$-algebra is a morphism of $A_\infty$- or $L_\infty$-algebras for which the chain map $\phi_1$ or $\psi_1$ is an isomorphism. Because $\phi_1$ or $\psi_1$ are always chain maps and thus descend to the cohomologies of the differentials $\sfm_1$ and $\mu_1$, we can extend the notion of quasi-isomorphism from ordinary chain complexes: a {\em quasi-isomorphism} is a morphism of $A_\infty$- or $L_\infty$-algebra for which $\phi_1$ or $\psi_1$ induces an isomorphism on the cohomology of the respective differential:
    \begin{equation}
        \fra \xrightarrow{~\phi~}\tilde \fra~,~~~\phi_1:H^\bullet_{\sfm_1}(\fra)\xrightarrow{~\cong~} H^\bullet_{\tilde \sfm_1}(\tilde \fra)\eand \frg \xrightarrow{~\psi~}\tilde \frg~,~~~\psi_1:H^\bullet_{\mu_1}(\frg)\xrightarrow{~\cong~} H^\bullet_{\tilde \mu_1}(\tilde \frg)~.
    \end{equation}
    
    It turns out that in essentially all situations, quasi-isomorphic $A_\infty$- or $L_\infty$-algebras should be regarded as equivalent. The $L_\infty$-algebras of two field theories are quasi-isomorphic if they have the same observables. That means in particular that they can be related by field redefinition, factoring out symmetries, integrating out fields, etc.

    \subsection{Structural theorems and field theories}
    
    There are a number of structural theorems that help us work with $L_\infty$-algebras and that also have concrete meaning for the discussion of field theories. First, let us define the following special classes of $A_\infty$- and $L_\infty$-algebras:
    \begin{itemize}
        \item[$\circ$] a {\em strict} $A_\infty$- or $L_\infty$-algebra is one in which the higher products $\sfm_i$ or $\mu_i$ vanish for $i>2$;
        \item[$\circ$] a {\em skeletal} $A_\infty$- or $L_\infty$-algebra is one in which the differential $\sfm_1$ or $\mu_1$ vanishes;
        \item[$\circ$] a {\em linearly contractible} $A_\infty$- or $L_\infty$-algebra is one in which the only non-trivial higher product is the differential $\sfm_1$ or $\mu_1$ and in which the corresponding cohomologies vanish.
    \end{itemize}
    Also, it is clear that we can construct the direct sums of homotopy algebras.
    Then we have the following structural theorems:
    \begin{itemize}
        \item[$\circ$] the {\em decomposition theorem}~\cite{Kajiura:0306332} states that any $A_\infty$- or $L_\infty$-algebra is isomorphic to the sum of a skeletal and a linearly contractible $A_\infty$- or $L_\infty$-algebra;
        \item[$\circ$] this directly implies the {\em minimal model theorem}~\cite{kadeishvili1982algebraic,Kajiura:0306332}, which states that any $A_\infty$- or $L_\infty$-algebra is quasi-isomorphic to a skeletal one, which is called a {\em minimal model} as it embeds into all quasi-isomorphic $L_\infty$-algebras;
        \item[$\circ$] the {\em strictification theorem}~\cite{igor1995,Berger:0512576} states that any $A_\infty$- or $L_\infty$-algebra is quasi-isomorphic to a strict one.
    \end{itemize}
    
    These theorems are now useful for the homological algebraic perspective on field theories. First, the strictification theorem implies that any field theory is equivalent to a field theory with only cubic interaction terms. That is easily seen for scalar field theories, where one can incorporate auxiliary fields with algebraic equations to render the action cubic, cf.~\cite{Macrelli:2019afx}. It is also well-known that Yang--Mills theory allows for a first-order formulation with cubic interaction vertices, cf.~the discussion in~\cite{Jurco:2018sby}. The abstract perspective makes this clear for any field theory.
    
    Second, the minimal model theorem tells us that for any field theory, there is an equivalent field theory with the same observables but without kinematical term (and thus without propagator). This equivalent field theory therefore has to encode the tree level scattering amplitudes of the original field theory. It is thus important to compute the minimal model of an $L_\infty$-algebra, which is best done with the homological perturbation lemma~\cite{Gugenheim1989:aa,gugenheim1991perturbation,Crainic:0403266}.
    
    \subsection{Minimal models from the homological perturbation lemma}
    
    Let us focus on the case of an $A_\infty$-algebra $\fra$; the discussion for an $L_\infty$-algebra is fully analogous, albeit slightly complicated by the graded symmetrisation operations that need to be included everywhere. The homological perturbation lemma (HPL) starts from the differential complex $(\fra,\sfm_1)$ underlying $\fra$ and the diagram 
    \begin{equation}\label{eq:initial_contraction}
        \begin{tikzcd}
            \ar[loop,out=160,in=200,distance=20,"\sfh" left] (\fra,\sfm_1)\arrow[r,twoheadrightarrow,shift left]{}{\sfp} & (\fra^\circ,0) \arrow[l,hookrightarrow,shift left]{}{\sfe}~,
        \end{tikzcd}
    \end{equation}
    where $\fra^\circ$ is the cohomology $H^\bullet_{\sfm_1}(\fra)$, $\sfp$ is the projection, $\sfe$ is an embedding of $\fra^\circ$ into $(\fra,\sfm_1)$, which requires a choice (i.e.~gauge fixing in the case of a gauge theory), and $\sfh$ is a contracting homotopy:
    \begin{equation}\label{eq:contracting_homotopy}
        \sfid=\sfm_1\circ\sfh+\sfh\circ \sfm_1+\sfe\circ\sfp~,~~~\sfp\circ\sfe=\sfid~.
    \end{equation}
    One can always redefine the maps $\sfe$, $\sfp$ and $\sfh$ such that the following relations are satisfied:
    \begin{equation}\label{eq:contractingBasic}
        \begin{gathered}
            \sfp\circ\sfh = \sfh\circ\sfe = \sfh\circ\sfh = 0~,~~~\sfp\circ \sfm_1 = \sfm_1\circ\sfe = 0~,
        \end{gathered}
    \end{equation}
    which simplifies the situation a bit more. The higher products $\sfm_i$ are then regarded as a perturbation of the differential $\sfm_1$ and the HPL gives the perturbations to the other maps so that a perturbed form of diagram~\eqref{eq:initial_contraction} is recovered.
    
    To simplify the discussion and to link it up to Feynman diagrams, we switch to the codifferential coalgebra picture. That is, we extend both $\sfp$ and $\sfe$ to coalgebra morphisms $\sfP_0$ and $\sfE_0$ between $\otimes^\bullet\fra$ and $\otimes^\bullet\fra^\circ$,
    \begin{subequations}
        \begin{equation}
            \sfP_0|_{\bigotimes^k\fra} \coloneqq \sfp^{\otimes^k}\eand
            \sfE_0|_{\bigotimes^k\fra^\circ} \coloneqq \sfe^{\otimes^k}~.
        \end{equation}
        The contracting homotopy $\sfh$ is extended to a map $\sfH_0:\bigotimes^\bullet\fra\to\bigotimes^\bullet\fra$ as follows:
        \begin{equation}\label{eq:contractingHomotopyZero}
            \begin{gathered}
                \sfH_0|_{\bigotimes^k\fra} \coloneqq \sum_{i+j=k-1}\sfid^{\otimes^i}\otimes\sfh\otimes(\sfe\circ\sfp)^{\otimes^j}~,
            \end{gathered}
        \end{equation}
    \end{subequations}
    which is sometimes called the ``tensor trick.'' We then arrive at a diagram
    \begin{equation}\label{eq:initial_contraction_H0}
        \begin{tikzcd}
            \ar[loop,out=160,in=200,distance=20,"\sfH_0" left] (\otimes^\bullet\fra,\sfD_0)\arrow[r,twoheadrightarrow,shift left]{}{\sfP_0} & (\otimes^\bullet\fra^\circ,0) \arrow[l,hookrightarrow,shift left]{}{\sfE_0}~,
        \end{tikzcd}
    \end{equation}
    where $\sfD_0$ is the continuation of $\sfm_1$ to a codifferential on $\otimes^\bullet\fra$ and equations~\eqref{eq:contractingBasic} induce the relations
    \begin{equation}\label{eq:contractingBasic_H0}
        \begin{gathered}
            \sfid = \sfD_0\circ\sfH_0+\sfH_0\circ \sfD_0+\sfE_0\circ\sfP_0~,\\
            \sfP_0\circ\sfE_0 = \sfid~,~~~
            \sfP_0\circ\sfH_0 = \sfH_0\circ\sfE_0 = \sfH_0\circ\sfH_0 = 0~,~~~
            \sfP_0\circ \sfD_0 = \sfD_0\circ\sfE_0 = 0~.
        \end{gathered}
    \end{equation}
    
    We now perturb $\sfD_0$ to $\sfD=\sfD_0+\sfD_\rint$, regarding $\sfD_\rint$ as small. The HPL then states that the corresponding deformation of~\eqref{eq:initial_contraction_H0} is given by the maps 
    \begin{equation}\label{eq:hpl_relations}
        \begin{aligned}
            \sfP &= \sfP_0\circ(1+\sfD_\rint\circ\sfH_0)^{-1},~~~&
            \sfH &= \sfH_0\circ(1+\sfD_\rint\circ\sfH_0)^{-1}~,\\
            \sfE &= (1+\sfH_0\circ\sfD_\rint)^{-1}\circ\sfE_0,~~~&
            \sfD^\circ &= \sfP\circ\sfD_\rint\circ\sfE_0~,
        \end{aligned}
    \end{equation}
    where $\sfD^\circ$ is now the codifferential on $\otimes^\bullet\fra^\circ$, encoding the higher products on the minimal model which we are after. By construction, $\sfE$ and $\sfP$ are compatible with the codifferentials,
    \begin{equation}
        \sfP\circ\sfD = \sfD^\circ\circ\sfP\eand\sfD\circ\sfE = \sfE\circ\sfD^\circ~.
    \end{equation}
    Later we use the fact that $\sfD^\circ$ is given by the morphism $\sfE$, which is constructed recursively:
    \begin{equation}\label{eq:recursion}
        \sfD^\circ=\sfP_0\circ\sfD_\rint\circ\sfE~,~~~\sfE=\sfE_0-\sfH_0\circ\sfD_\rint\circ\sfE~.
    \end{equation}
    This recursion relation is responsible for recursion relations in computations of scattering amplitudes.
    
    \section{Scattering amplitudes}
    
    The idea of perturbative quantum field theory is quite literally implemented in the homological perturbation lemma. Given an $A_\infty$-algebra $\fra$ for a Lagrangian field theory, the space $\fra_1$ is the set of fields while the space $\fra_1^\circ$ is the set of free on-shell fields. The recursion relation~\eqref{eq:recursion} encodes precisely the tree level Feynman diagram expansion. A slight extension of the homological perturbation lemma incorporates loops and yet again, useful recursion relations are obtained.
    
    \subsection{Tree level}
    
    We start with the example of scalar field theory, using the $A_\infty$-algebra $\fra$ defined in~\eqref{eq:a_infty_scalar}. We would like to compute the tree level four-point amplitude. This is given by the higher products in the minimal model of the $L_\infty$-algebra corresponding to $\fra$ or to the symmetrised expression
    \begin{equation}\label{eq:amplitude_scalar}
        \begin{aligned}
            \caA(\varphi_1,\varphi_2,\varphi_3,\varphi_4)&=\sum_{\sigma\in S_4/\IZ_4}\langle \varphi_{\sigma(4)},\sfm^\circ_3(\varphi_{\sigma(1)},\varphi_{\sigma(2)},\varphi_{\sigma(3)})\rangle\\
            &=\sum_{\sigma\in S_3}\langle \varphi_4,\sfm^\circ_3(\varphi_{\sigma(1)},\varphi_{\sigma(2)},\varphi_{\sigma(3)})\rangle~.
        \end{aligned}
    \end{equation}
    The higher product $\sfm_3^\circ$ appears now in the codifferential $\sfD^\circ$, and $\sfm_3^\circ$ is the restriction of $\sfD^\circ$ to the domain $\otimes^3\fra$, with subsequent projection onto $\fra$. Let us introduce simplifying notation. For any operator $\caO:\otimes^\bullet \fra_1\rightarrow \otimes^\bullet \fra_2$, we denote the restriction and projection onto $j$ inputs and $i$ outputs by
    \begin{equation}\label{eq:notation_projection}
        \caO^{i,j} \coloneqq \sfpr_{\otimes^i\fra_2}\circ\caO\big|_{\otimes^j\fra_1}~.
    \end{equation}
    In particular, $\sfm_3=\sfD^{\circ\,1,3}$. The recursion relation~\eqref{eq:recursion} translates to the explicit recursion
    \begin{equation}\label{eq:recursionRelation2}
        \begin{aligned}
            \sfD^{\circ\,1,3}&=\sfP_0^{1,1}\circ (\sfD^{1,2}_\rint\circ \sfE^{2,3}+\sfD^{1,3}_\rint\circ \sfE^{3,3})~,\\
            \sfE^{i,j}&=\delta^{ij}\sfE^{i,i}_0-\sfH_0\circ\sum_{k=2}^{i+2}\sfD_\rint^{i,k}\circ\sfE^{k,j}~,
        \end{aligned}
    \end{equation}
    so that
    \begin{equation}
        \sfD^{\circ\,1,3}=\sfP_0^{1,1}\circ \big(\sfD^{1,2}_\rint\circ \sfH_0\circ \sfD^{2,3}_\rint+\sfD^{1,3}_\rint\big)\circ\sfE^{3,3}_0~.
    \end{equation}
    Explicitly, we compute
    \begin{equation}\label{eq:m_circ_3}
        \begin{aligned}
            \sfm^\circ_3(\varphi_1,\varphi_2,\varphi_3)&=\sfD^{\circ\,1,3}(\varphi_1\otimes \varphi_2\otimes \varphi_3)\\
            &=\Big(\sfP_0^{1,1}\circ \big(\sfD^{1,2}_\rint\circ \sfH_0\circ \sfD^{2,3}_\rint+\sfD^{1,3}_\rint\big)\circ\sfE^{3,3}_0\Big)(\varphi_1\otimes \varphi_2\otimes \varphi_3)\\
            &=\Big(\sfP_0^{1,1}\circ \big(\sfD^{1,2}_\rint\circ \sfH_0\circ \sfD^{2,3}_\rint+\sfD^{1,3}_\rint\big)\Big)(\sfe(\varphi_1)\otimes \sfe(\varphi_2)\otimes \sfe(\varphi_3))\\
            &=\sfp\Big(\sfm_2(\sfh(\sfm_2(\sfe(\varphi_1),\sfe(\varphi_2))),\sfe(\varphi_3))+\sfm_2(\sfe(\varphi_1),\sfh(\sfm_2(\sfe(\varphi_2),\sfe(\varphi_3))))+\\
            &\hspace{3cm}+ \sfm_3(\sfe(\varphi_1),\sfe(\varphi_2),\sfe(\varphi_3))\Big)~.
        \end{aligned}
    \end{equation}
    Here, $\sfe$ embeds the asymptotic on-shell fields $\varphi_i$ into the full set of interacting fields. The relation~\eqref{eq:contracting_homotopy} shows that $\sfh$ is the inverse to the kinematical operator $\sfm_1$ on field space with the free fields removed, i.e.~the propagator. The maps $\sfm_2$ and $\sfm_3$ correspond to cubic and quartic interaction vertices. We can thus translate the expression~\eqref{eq:m_circ_3} into Feynman diagrams for a ``current'' (three incoming fields, $\varphi_1$, $\varphi_2$, and $\varphi_3$, and one outgoing antifield, $\varphi_4^+$) and plugging this into formula~\eqref{eq:amplitude_scalar}, we recover the expected four-point tree level Feynman diagrams:
    \begin{equation}
        \vcenter{\hbox{
                \begin{axopicture}{(80,80)(0,0)}
                    \SetArrowStroke{0.5}
                    \Vertex(55,40){2}
                    \Vertex(25,40){2}
                    \Line(25,40)(55,40)
                    \Line(10,10)(25,40)
                    \Line(10,70)(25,40)
                    \Line(70,10)(55,40) 
                    \Line(70,70)(55,40) 
                    \Text(5,5){$\varphi_1$}
                    \Text(75,5){$\varphi_2$}
                    \Text(5,75){$\varphi_4$}
                    \Text(75,75){$\varphi_3$}
                    \Text(15,40){$\kappa$}
                    \Text(65,40){$\kappa$}
                \end{axopicture}
                ~~~
                \begin{axopicture}{(80,80)(0,0)}
                    \SetArrowStroke{0.5}
                    \Vertex(40,55){2}
                    \Vertex(40,25){2}
                    \Line(40,25)(40,55)
                    \Line(10,10)(40,25)
                    \Line(10,70)(40,55)
                    \Line(70,10)(40,55) 
                    \Line(70,70)(40,25) 
                    \Text(5,5){$\varphi_1$}
                    \Text(75,5){$\varphi_2$}
                    \Text(5,75){$\varphi_4$}
                    \Text(75,75){$\varphi_3$}
                    \Text(40,15){$\kappa$}
                    \Text(40,65){$\kappa$}
                \end{axopicture}
                ~~~
                \begin{axopicture}{(80,80)(0,0)}
                    \SetArrowStroke{0.5}
                    \Vertex(40,55){2}
                    \Vertex(40,25){2}
                    \Line(40,25)(40,55)
                    \Line(10,10)(40,25)
                    \Line(10,70)(40,55)
                    \Line(70,10)(40,25) 
                    \Line(70,70)(40,55) 
                    \Text(5,5){$\varphi_1$}
                    \Text(75,5){$\varphi_2$}
                    \Text(5,75){$\varphi_4$}
                    \Text(75,75){$\varphi_3$}
                    \Text(40,15){$\kappa$}
                    \Text(40,65){$\kappa$}
                \end{axopicture}
                ~~~
                \begin{axopicture}{(80,80)(0,0)}
                    \SetArrowStroke{0.5}
                    \Vertex(40,40){2}
                    \Line(10,10)(40,40)
                    \Line(10,70)(40,40)
                    \Line(70,10)(40,40) 
                    \Line(70,70)(40,40) 
                    \Text(5,5){$\varphi_1$}
                    \Text(75,5){$\varphi_2$}
                    \Text(5,75){$\varphi_4$}
                    \Text(75,75){$\varphi_3$}
                    \Text(40,50){$\lambda$}
                \end{axopicture}
            }}
    \end{equation}
        
    A more precise analysis requires us to restrict the field space $C^\infty(\IR^{1,3})$ we used in the graded vector space underlying $\fra$ to a sum of interacting and on-shell fields represented by suitable functions on $\IR^{1,3}$. This discussion can be found in~\cite{Macrelli:2019afx}, but we suppress these details here and in the following, as they are mostly technical. 
    
    The same procedure can be applied to Yang--Mills theory. Here, the $A_\infty$-algebra~\eqref{eq:a_infty_YM} factorises into the tensor product of a kinematical $A_\infty$-algebra and a colour Lie algebra~\cite{Jurco:2019yfd}. Moreover, the recursion relation~\eqref{eq:recursion} for $\sfE$ induces a recursion relation for currents, which is known as the Berends--Giele recursion relation~\cite{Berends:1987me}, which was used in the same paper to prove the Parke--Taylor formula conjectured in~\cite{Parke:1986gb}.
    
    \subsection{Loop level}
    
    Let us return to the homological perturbation lemma. We note that the BV formalism gives us a guideline as to how to incorporate the quantum effects and to go to full loop level. Namely, the BV differential $Q_{\rm BV}$ should be replaced by a linear combination of the BV differential with the BV Laplacian $\Delta$, with a transition from classical to quantum master equation:
    \begin{equation}\label{eq:BV_classical_to_quantum}
        Q_{\rm BV} \coloneqq \{S_{\rm BV},-\}~,~~~\{S_{\rm BV},S_{\rm BV}\}=0~~~\longrightarrow~~~\hbar \Delta+\{S_{\rm BV},-\}~,~~~2\hbar \Delta S_{\rm BV}+\{S_{\rm BV},S_{\rm BV}\}=0~.
    \end{equation}
    In some cases, the quantum master equation requires corrections to the classical $S_{\rm BV}$ in powers of $\hbar$. For (unregularised) scalar field theory and Yang--Mills theory, however, the classical BV action also satisfies the quantum master equation.
    
    We have to translate the replacement~\eqref{eq:BV_classical_to_quantum} to the dual, codifferential coalgebra picture that we have been using at tree level. Recall that the BV Laplacian is a product of two functional derivatives, 
    \begin{equation}
        \Delta=\frac{\partial^2}{\partial \phi^A\partial\phi^+_A}~,
    \end{equation}
    one with respect to a field $\phi^A$ and one with respect to the corresponding antifield $\phi^+_A$ with identical labels $A$ (momentum, polarisation, colour factor, etc.). Dually, $\Delta^*$ should insert pairs of fields and antifields everywhere into the tensor product. In the case of a scalar field theory with fields $\varphi_{1,2}\in \fra_1$, for example, we expect
    \begin{equation}\label{eq:actionDualBVLScalar}
        \begin{aligned}
            \Delta^*(\varphi_1\otimes\varphi_2)\ &=\ \int\frac{\rmd^4k}{(2\pi)^4}\Big\{\psi(k)\otimes\psi^+(k)\otimes\varphi_1\otimes\varphi_2+\psi(k)\otimes\varphi_1\otimes\psi^+(k)\otimes\varphi_2+\cdots\\
            &\kern2cm+\psi^+(k)\otimes\psi(k)\otimes\varphi_1\otimes\varphi_2+\psi^+(k)\otimes\varphi_1\otimes\psi(k)\otimes\varphi_2+\cdots\Big\}~,
        \end{aligned}
    \end{equation}
    where $\psi(k)$ is a (momentum space) basis of the field space $\fra_1$ and $\psi^+(k)$ of the antifield space $\fra_2$. A more precise analysis using the actual field space shows that in the operator $\sfH_0\circ (\rmi \hbar \Delta^*)$, only the first line of~\eqref{eq:actionDualBVLScalar} is non-vanishing, while the second line is canceled by the maps in $\sfH$. 
    
    The transition from classical to quantum master equation~\eqref{eq:BV_classical_to_quantum} corresponds then to a replacement
    \begin{equation}\label{eq:substitution}
        \sfD_\rint\ \to\ \sfD_\rint-\rmi\hbar\Delta^*~,
    \end{equation}
    and we can still treat $\sfD_\rint-\rmi\hbar\Delta^*$ as a perturbation and thus apply the homological perturbation lemma. The recursion relation~\eqref{eq:recursion} then takes the form
    \begin{equation}\label{eq:def_D_circ}
        \sfD^\circ=\sfP_0\circ(\sfD_\rint-\rmi\hbar\Delta^*)\circ\sfE~,~~~~~
        \sfE=\sfE_0-\sfH_0\circ(\sfD_\rint-\rmi\hbar\Delta^*)\circ\sfE~.
    \end{equation}
    Note that the map $\sfD_\rint-\rmi\hbar\Delta^*$ is no longer a coderivation as it fails to satisfy the co-Leibniz rule~\eqref{eq:coLeibniz}. As a result, the maps $\sfD^\circ$ and $\sfE$ are no longer coalgebra morphisms, in general, and the structure encoded by $\sfD^\circ$ is not a classical $A_\infty$-algebra, but a {\em quantum $A_\infty$-algebra}. In a quantum homotopy algebra, the homotopy relations are corrected by orders in $\hbar$, cf.~\cite{Markl:1997bj} for details on quantum $L_\infty$-algebras. It is this quantum minimal model that describes the full Feynman diagram expansion.
    
    Just as at tree level, we also have a finite recursion relation at loop level~\cite{Jurco:2019yfd}. Using again the notation~\eqref{eq:notation_projection}, the morphism $\sfE^{i,j}$, restricted to $\ell$ loops and $v$ interaction vertices satisfies
    \begin{equation}\label{eq:recursionRelation_loop}
        \begin{aligned}
            \sfE^{i,j}_{\ell,v}=\delta_\ell^0\delta_v^0\delta^{ij}\sfE^{ii}_0-\sfH_0\circ\sum_{k=2}^{i+2}\sfD_\rint^{i,k}\circ\sfE^{k,j}_{\ell,v-1}+\rmi\hbar\,\sfH_0\circ\Delta^*\circ\sfE^{i-2,j}_{\ell-1,v}~,
        \end{aligned}
    \end{equation}
    where $\sfE^{i,j}_{\ell,v}=0$ for $\ell<0$ or $v<0$. This formula can directly be inserted into a computer algebra programme to generate the perturbative expansion to any desired order.
    
    As an application, let us briefly examine the 1-loop correction to the 2-point function in scalar field theory with $A_\infty$-algebra~\eqref{eq:a_infty_scalar}. That is, we compute $\sfD^{\circ\,1,1}$ up to one loop. Note that~\eqref{eq:def_D_circ} and~\eqref{eq:recursionRelation_loop} simplify to 
    \begin{equation}\label{eq:1-loop-2-point}
        \begin{aligned}
            \sfD^{\circ\,1,1}&=\sfP^{1,1}_0\circ (\sfD^{1,3}_\rint\circ \sfE^{3,1}_{1,0}+\sfD^{1,2}_\rint\circ \sfE^{2,1}_{1,1})\\
            &=\sfP^{1,1}_0\circ (\sfD^{1,3}_\rint\circ \rmi\hbar\Delta^*\circ \sfE_0^{1,1}+\sfD^{1,2}_\rint\circ \sfH_0\circ \sfD^{2,3}\circ \rmi\hbar\Delta^*\circ \sfE_0^{1,1})~.
        \end{aligned}
    \end{equation}
    We now apply this expression to a field $\varphi\in \fra^\circ_1$. First, we note that 
    \begin{equation}
        \begin{aligned}
            (\Delta^*\circ \sfE_0^{1,1})(\varphi)&=\int\frac{\rmd^4k}{(2\pi)^4}\Big\{\psi(k)\otimes\psi^+(k)\otimes\sfe(\varphi)+\psi(k)\otimes\sfe(\varphi)\otimes\psi^+(k)+\\
            &\hspace{2cm}+\sfe(\varphi)\otimes\psi(k)\otimes\psi^+(k)+
            \psi^+(k)\otimes\psi(k)\otimes\sfe(\varphi)+\\
            &\hspace{2cm}+\psi^+(k)\otimes\sfe(\varphi)\otimes\psi(k)+\sfe(\varphi)\otimes\psi^+(k)\otimes\psi(k)\Big\}~,
        \end{aligned}
    \end{equation}
    and the full formula~\eqref{eq:1-loop-2-point} is best depicted as the following diagrams:
    
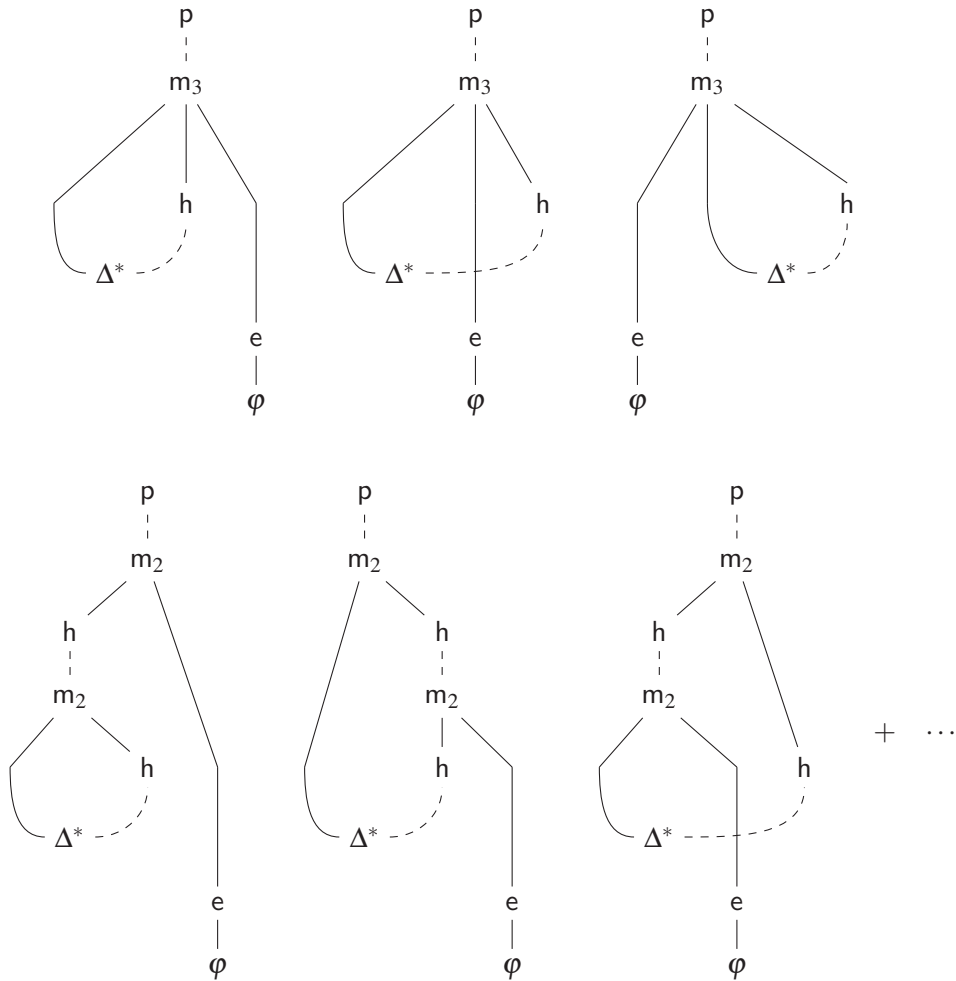
\begin{equation*}
        \begin{tikzpicture}[baseline={([yshift=-.5ex]current bounding box.center)}]
            \matrix (m) [matrix of nodes, ampersand replacement=\&, column sep = 0.3cm, row sep = 0.4cm]{
                {} \& {} \& $\sfp$ \& {}  \\
                {} \& {} \& $\sfm_3$ \& {}  \\
                {} \& {} \& {} \& {}  \\
                {} \& {} \& $\sfh$ \& {}  \\
                {} \& $\Delta^*$ \& {} \& {}  \\
                {} \& {} \& {} \& $\sfe$ \\
                {} \& {} \& {} \& $\varphi$ \\
            };
            \draw[dashed] (m-1-3) -- (m-2-3) ;
            \draw (m-4-4.north) -- (m-2-3) ;
            \draw (m-4-3) -- (m-2-3) ;
            \draw (m-4-1.north) -- (m-2-3) ;
            \draw (m-4-4.north) -- (m-6-4.north) ;
            \draw (m-5-2.west) to[out=-180,in=-90] (m-4-1.north);
            \draw[dashed] (m-5-2.east) to[out=0,in=-90] (m-4-3.south);
            \draw (m-6-4.south) -- (m-7-4.north) ;
        \end{tikzpicture}
        \hspace{0.5cm}
        \begin{tikzpicture}[baseline={([yshift=-.5ex]current bounding box.center)}]
            \matrix (m) [matrix of nodes, ampersand replacement=\&, column sep = 0.3cm, row sep = 0.4cm]{
                {} \& {} \& $\sfp$ \& {}  \\
                {} \& {} \& $\sfm_3$ \& {}  \\
                {} \& {} \& {} \& {}  \\
                {} \& {} \& {} \& $\sfh$  \\
                {} \& $\Delta^*$ \& {} \& {}  \\
                {} \& {} \& $\sfe$ \& {} \\
                {} \& {} \& $\varphi$ \& {} \\
            };
            \draw[dashed] (m-1-3) -- (m-2-3) ;
            \draw (m-4-3.north) -- (m-2-3) ;
            \draw (m-4-4) -- (m-2-3) ;
            \draw (m-4-1.north) -- (m-2-3) ;
            \draw (m-4-3.north) -- (m-6-3.north) ;
            \draw (m-5-2.west) to[out=-180,in=-90] (m-4-1.north);
            \draw[dashed] (m-5-2.east) to[out=0,in=-90] (m-4-4.south);
            \draw (m-6-3.south) -- (m-7-3.north) ;
        \end{tikzpicture}
        \hspace{0.5cm}
        \begin{tikzpicture}[baseline={([yshift=-.5ex]current bounding box.center)}]
            \matrix (m) [matrix of nodes, ampersand replacement=\&, column sep = 0.3cm, row sep = 0.4cm]{
                {} \& $\sfp$ \& {} \& {}  \\
                {} \& $\sfm_3$ \& {} \& {}  \\
                {} \& {} \& {} \& {}  \\
                {} \& {} \& {} \& $\sfh$  \\
                {} \& {} \& $\Delta^*$ \& {}  \\
                $\sfe$ \& {} \& {} \& {} \\
                $\varphi$ \& {} \& {} \& {} \\
            };
            \draw[dashed] (m-1-2) -- (m-2-2) ;
            \draw (m-4-1.north) -- (m-2-2) ;
            \draw (m-4-2) -- (m-2-2) ;
            \draw (m-4-4.north) -- (m-2-2) ;
            \draw (m-4-1.north) -- (m-6-1.north) ;
            \draw (m-5-3.west) to[out=-180,in=-90] (m-4-2.north);
            \draw[dashed] (m-5-3.east) to[out=0,in=-90] (m-4-4.south);
            \draw (m-6-1.south) -- (m-7-1.north) ;
        \end{tikzpicture}\hspace{0.5cm}
    \end{equation*}
    \begin{equation*}
        \begin{tikzpicture}[baseline={([yshift=-.5ex]current bounding box.center)}]
            \matrix (m) [matrix of nodes, ampersand replacement=\&, column sep = 0.3cm, row sep = 0.4cm]{
                {} \& {} \& $\sfp$ \& {}  \\
                {} \& {} \& $\sfm_2$ \& {}  \\
                {} \& $\sfh$ \& {} \& {}  \\
                {} \& $\sfm_2$ \& {} \& {}  \\
                {} \& {} \& $\sfh$ \& {}  \\
                {} \& $\Delta^*$ \& {} \& {}  \\
                {} \& {} \& {} \& $\sfe$ \\
                {} \& {} \& {} \& $\varphi$ \\
            };
            \draw[dashed] (m-1-3) -- (m-2-3) ;
            \draw (m-5-4.north) -- (m-2-3) ;
            \draw (m-5-3) -- (m-4-2) ;
            \draw (m-2-3) -- (m-3-2) ;
            \draw[dashed] (m-3-2) -- (m-4-2) ;
            \draw (m-5-1.north) -- (m-4-2) ;
            \draw (m-5-4.north) -- (m-7-4.north) ;
            \draw (m-6-2.west) to[out=-180,in=-90] (m-5-1.north);
            \draw[dashed] (m-6-2.east) to[out=0,in=-90] (m-5-3.south);
            \draw (m-7-4.south) -- (m-8-4.north) ;
        \end{tikzpicture}
        \hspace{0.5cm}
        \begin{tikzpicture}[baseline={([yshift=-.5ex]current bounding box.center)}]
            \matrix (m) [matrix of nodes, ampersand replacement=\&, column sep = 0.3cm, row sep = 0.4cm]{
                {} \& $\sfp$ \& {} \& {}  \\
                {} \& $\sfm_2$ \& {} \& {}  \\
                {} \& {} \& $\sfh$ \& {}  \\
                {} \& {} \& $\sfm_2$ \& {}  \\
                {} \& {} \& $\sfh$ \& {}  \\
                {} \& $\Delta^*$ \& {} \& {}  \\
                {} \& {} \& {} \& $\sfe$ \\
                {} \& {} \& {} \& $\varphi$ \\
            };
            \draw[dashed] (m-3-3) -- (m-4-3) ;
            \draw[dashed] (m-1-2) -- (m-2-2) ;
            \draw (m-5-4.north) -- (m-4-3) ;
            \draw (m-5-3) -- (m-4-3) ;
            \draw (m-2-2) -- (m-3-3) ;
            \draw (m-5-1.north) -- (m-2-2) ;
            \draw (m-5-4.north) -- (m-7-4.north) ;
            \draw (m-6-2.west) to[out=-180,in=-90] (m-5-1.north);
            \draw[dashed] (m-6-2.east) to[out=0,in=-90] (m-5-3.south);
            \draw (m-7-4.south) -- (m-8-4.north) ;
        \end{tikzpicture}
        \hspace{0.5cm}
        \begin{tikzpicture}[baseline={([yshift=-.5ex]current bounding box.center)}]
            \matrix (m) [matrix of nodes, ampersand replacement=\&, column sep = 0.3cm, row sep = 0.4cm]{
                {} \& {} \& $\sfp$ \& {}  \\
                {} \& {} \& $\sfm_2$ \& {}  \\
                {} \& $\sfh$ \& {} \& {}  \\
                {} \& $\sfm_2$ \& {} \& {}  \\
                {} \& {} \& {} \& $\sfh$  \\
                {} \& $\Delta^*$ \& {} \& {}  \\
                {} \& {} \& $\sfe$ \& {} \\
                {} \& {} \& $\varphi$ \& {} \\
            };
            \draw[dashed] (m-3-2) -- (m-4-2) ;
            \draw[dashed] (m-1-3) -- (m-2-3) ;
            \draw (m-5-3.north) -- (m-4-2) ;
            \draw (m-5-4) -- (m-2-3) ;
            \draw (m-5-1.north) -- (m-4-2) ;
            \draw (m-3-2) -- (m-2-3) ;
            \draw (m-5-3.north) -- (m-7-3.north) ;
            \draw (m-6-2.west) to[out=-180,in=-90] (m-5-1.north);
            \draw[dashed] (m-6-2.east) to[out=0,in=-90] (m-5-4.south);
            \draw (m-7-3.south) -- (m-8-3.north) ;
        \end{tikzpicture}\hspace{0.5cm}+~~~\cdots
    \end{equation*}
    Here, $\varphi$ denotes the incoming on-shell field, the map $\sfe$ embeds it into the set of the full interacting fields, dashed lines denote antifields, which are produced by the higher products $\sfm_{2,3}:\fra_1^{\times 2,3}\rightarrow \fra_2$ and the dual BV Laplacian $\Delta^*$, and $\sfp$ is the projection of an antifield back onto the on-shell antifields. In the second line, a more precise analysis using  the actual field space shows that the third diagram vanishes. Moreover, the ellipsis in the second line refer to another three diagrams. We note that strings of operations as the summands in~\eqref{eq:1-loop-2-point} encode several Feynman diagrams.
    
    The 2-point amplitude at one loop is given by
    \begin{equation}
        \caA^{\rm 1~loop}_2(\varphi_1,\varphi_2)=\langle~\varphi_1~,~\sfD^{\circ\,1,1}(\varphi_2)~\rangle~.
    \end{equation}
    and the above diagrams turn indeed into the expected Feynman diagrams with the expected symmetry factors.

    \subsection{Applications}
    
    In this last section, let us briefly point out some applications of the homological algebraic perspective on perturbative quantum field theory. So far, we merely saw that any quantum field theory can be recast in an equivalent quantum field theory with cubic interaction vertices and that the Berends--Giele tree level recursion relations for Yang--Mills currents generalise to loop level and to an arbitrary Lagrangian field theory.
    
    One of the key reasons to consider $A_\infty$-algebras over $L_\infty$-algebras is that they allow for an easy analysis of the colour structure of loop amplitudes in Yang--Mills theory. Recall that a Yang--Mills field can either be regarded as a one-form taking values in the vector space underlying the gauge Lie algebra or, in the case of a matrix gauge algebra, as a matrix-valued one-form. In the second case, Feynman diagrams ``thicken'' to ribbon graphs and each line is replaced by a pair of lines, one for each matrix index. The Lie bracket is replaced by the associative matrix product, and this is best encoded in an $A_\infty$-algebra. Following the contractions of indices in loop diagrams, which is particularly easy in our formalism, it becomes clear that Feynman diagrams split into planar and non-planar diagrams and that planar diagrams dominate non-planar ones by at least one factor $N$, where $N$ is the rank of the gauge group. A more precise formula is readily derived~\cite{Jurco:2019yfd} in our picture. Moreover, the one-loop amplitude of Yang--Mills theory is fully determined by its planar part. This has been known for some time~\cite{Bern:1994zx}, but the proof of this relation is simplified in our formalism~\cite{Jurco:2019yfd}.
    
    In general, our formalism is well-suited to prove combinatorial identities. As a new example, let us give a short proof that the tree level Berends--Giele currents in Yang--Mills theory vanish when summed over all shuffles; a traditional proof is found already in~\cite{Berends:1988zn}, see also~\cite{Kol:1403.6837} or~\cite[Appendix A]{Lee:2015upy}. We start from the colour-stripped form of the Yang--Mills $A_\infty$-algebra~\eqref{eq:a_infty_YM}, cf.~\cite{Jurco:2019yfd}, which has in particular higher products\footnote{denoted by $m_i$ to distinguish from the full higher products $\sfm_i$}
    \begin{equation}
        \begin{aligned}
            m_2(A_1,A_2)&=\kappa \big(\rmd^\dagger(A_1\wedge A_2)+\star (A_1\wedge \star\rmd A_2)-\star((\star\rmd A_1)\wedge A_2)\big)~,\\
            m_3(A_1,A_2,A_3)&=\kappa^2\big(\star(A_1\wedge\star(A_2\wedge A_3))}-{\star(\star(A_1\wedge A_2)\wedge A_3)\big)~,
        \end{aligned}
    \end{equation}
    where here the $A_i$ are plain differential forms. Consider now the colour-stripped current 
    \begin{equation}
        \sfE^{\circ 1,n}(A_1,\dots,A_n)~,
    \end{equation}
    and partition the input fields $A_i$ into two pairwise disjoint, non-empty subsets $\Phi_1=(A_1,\ldots,A_m)$ and $\Phi_2=(A_{m+1},\dots,A_n)$. Then, at tree level,
    \begin{equation}\label{eq:BG_identity}
        \sum_{\sigma\in {\rm Sh}(m;n)}\sfE^{\circ\,1,n}(A_{\sigma^{-1}(1)}\otimes\dots\otimes A_{\sigma^{-1}(n)})=0~,
    \end{equation}
    where the sum runs over all shuffles, i.e.~permutations preserving the relative order of the $A_i$ in $\Phi_1$ and $\Phi_2$. The map $\sfE^{1,n}$ can be depicted as a sum of trees with binary and ternary nodes, whose leaves are the sequence of input fields $A_{\sigma^{-1}(1)}\otimes\dots\otimes A_{\sigma^{-1}(n)}$. Each such tree will contain maximal subtrees where all the leaves are either from $\Phi_1$ or $\Phi_2$, which we call {\em pure subtrees} of type $\Phi_1$ or $\Phi_2$. Two or three of these pure subtrees are then joined together by nodes $m_2$ or $m_3$, originating from actions of $\sfD_\rint$. There are now eight distinct such ordered joins, depicted below. If one of a diagrams appears in a shuffle sum, then so do the others in the same row. Moreover, due to the symmetry properties of $m_2$ and $m_3$, these sums vanish:
    \begin{subequations}
        \begin{equation}
            \begin{tikzpicture}[baseline={([yshift=-.5ex]current bounding box.center)}]
                \matrix (m) [matrix of nodes, ampersand replacement=\&, column sep = 0.1cm, row sep = 0.4cm]{
                    {} \& {} \& {}  \\
                    {} \& $m_2$ \& {}  \\
                    $\Phi_1$ \& {} \& $\Phi_2$ \\
                };
                \draw[dashed] (m-1-2) -- (m-2-2) ;
                \draw (m-2-2) -- (m-3-1) ;
                \draw (m-2-2) -- (m-3-3) ;
            \end{tikzpicture}
            +
            \begin{tikzpicture}[baseline={([yshift=-.5ex]current bounding box.center)}]
                \matrix (m) [matrix of nodes, ampersand replacement=\&, column sep = 0.1cm, row sep = 0.4cm]{
                    {} \& {} \&  {}  \\
                    {} \& $m_2$ \& {}  \\
                    $\Phi_2$ \& {} \& $\Phi_1$  \\
                };
                \draw[dashed] (m-1-2) -- (m-2-2) ;
                \draw (m-2-2) -- (m-3-1) ;
                \draw (m-2-2) -- (m-3-3) ;
            \end{tikzpicture}=0
        \end{equation}
        \begin{equation}
            \begin{tikzpicture}[baseline={([yshift=-.5ex]current bounding box.center)}]
                \matrix (m) [matrix of nodes, ampersand replacement=\&, column sep = 0.1cm, row sep = 0.4cm]{
                    {} \& {} \&  {}  \\
                    {} \& $m_3$ \& {}  \\
                    $\Phi_1$ \& $\tilde \Phi_1$ \& $\Phi_2$  \\
                };
                \draw[dashed] (m-1-2) -- (m-2-2) ;
                \draw (m-2-2) -- (m-3-1) ;
                \draw (m-2-2) -- (m-3-2) ;
                \draw (m-2-2) -- (m-3-3) ;
            \end{tikzpicture}
            +
            \begin{tikzpicture}[baseline={([yshift=-.5ex]current bounding box.center)}]
                \matrix (m) [matrix of nodes, ampersand replacement=\&, column sep = 0.1cm, row sep = 0.4cm]{
                    {} \& {} \&  {}  \\
                    {} \& $m_3$ \& {}  \\
                    $\Phi_1$ \& $\Phi_2$ \& $\tilde \Phi_1$  \\
                };
                \draw[dashed] (m-1-2) -- (m-2-2) ;
                \draw (m-2-2) -- (m-3-1) ;
                \draw (m-2-2) -- (m-3-2) ;
                \draw (m-2-2) -- (m-3-3) ;
            \end{tikzpicture}
            +
            \begin{tikzpicture}[baseline={([yshift=-.5ex]current bounding box.center)}]
                \matrix (m) [matrix of nodes, ampersand replacement=\&, column sep = 0.1cm, row sep = 0.4cm]{
                    {} \& {} \&  {}  \\
                    {} \& $m_3$ \& {}  \\
                    $\Phi_2$ \& $\Phi_1$ \& $\tilde \Phi_1$  \\
                };
                \draw[dashed] (m-1-2) -- (m-2-2) ;
                \draw (m-2-2) -- (m-3-1) ;
                \draw (m-2-2) -- (m-3-2) ;
                \draw (m-2-2) -- (m-3-3) ;
            \end{tikzpicture}
            =0
        \end{equation}
        \begin{equation}
            \begin{tikzpicture}[baseline={([yshift=-.5ex]current bounding box.center)}]
                \matrix (m) [matrix of nodes, ampersand replacement=\&, column sep = 0.1cm, row sep = 0.4cm]{
                    {} \& {} \&  {}  \\
                    {} \& $m_3$ \& {}  \\
                    $\Phi_1$ \& $\Phi_2$ \& $\tilde \Phi_2$  \\
                };
                \draw[dashed] (m-1-2) -- (m-2-2) ;
                \draw (m-2-2) -- (m-3-1) ;
                \draw (m-2-2) -- (m-3-2) ;
                \draw (m-2-2) -- (m-3-3) ;
            \end{tikzpicture}
            +
            \begin{tikzpicture}[baseline={([yshift=-.5ex]current bounding box.center)}]
                \matrix (m) [matrix of nodes, ampersand replacement=\&, column sep = 0.1cm, row sep = 0.4cm]{
                    {} \& {} \&  {}  \\
                    {} \& $m_3$ \& {}  \\
                    $\Phi_2$ \& $\tilde \Phi_2$ \& $\Phi_1$  \\
                };
                \draw[dashed] (m-1-2) -- (m-2-2) ;
                \draw (m-2-2) -- (m-3-1) ;
                \draw (m-2-2) -- (m-3-2) ;
                \draw (m-2-2) -- (m-3-3) ;
            \end{tikzpicture}
            +
            \begin{tikzpicture}[baseline={([yshift=-.5ex]current bounding box.center)}]
                \matrix (m) [matrix of nodes, ampersand replacement=\&, column sep = 0.1cm, row sep = 0.4cm]{
                    {} \& {} \&  {}  \\
                    {} \& $m_3$ \& {}  \\
                    $\Phi_2$ \& $\Phi_1$ \& $\tilde \Phi_2$  \\
                };
                \draw[dashed] (m-1-2) -- (m-2-2) ;
                \draw (m-2-2) -- (m-3-1) ;
                \draw (m-2-2) -- (m-3-2) ;
                \draw (m-2-2) -- (m-3-3) ;
            \end{tikzpicture}
            =0
        \end{equation}
    \end{subequations}
    Here $\Phi_1,\tilde \Phi_1$ and $\Phi_2,\tilde \Phi_2$ depict pure trees of type $\Phi_1$ and $\Phi_2$, respectively. We thus conclude~\eqref{eq:BG_identity}, which, as pointed out in~\cite{Lee:2015upy}, implies the Kleiss--Kuijf relations~\cite{Kleiss:1988ne,DelDuca:1999rs}.

    \subsection{Outlook}
    
    An obvious problem to attack form the homological algebraic perspective is certainly the derivation of the BCJ double copy formulas~\cite{Bern:2008qj,Bern:2010ue} which state that tree level amplitudes in gravity can be obtained from two copies of colour-stripped Yang--Mills tree level amplitudes. Since the origin of this relation lies in the open/closed duality of strings and because the underlying string field theories are best described using homotopy algebras, our formalism seems to be the natural framework to attack this problem.
    
    Finally, we saw that physical equivalence of classical field theories translates to quasi-isomor\-phisms of their corresponding homotopy algebras. It would be very interesting to analyse the corresponding statement for the minimal models of quantum homotopy algebras. In this context, the renormalisation group flow should be a quasi-isomorphism of quantum homotopy algebras.
    
\bibliographystyle{latexeu}
\bibliography{bigone}

\end{document}